\documentclass[journal]{IEEEtran}

\usepackage[latin9]{inputenc}
\usepackage{amssymb}
\usepackage{amsthm}
\newtheorem{theorem}{Theorem}[section]

\newtheorem{lemma}[theorem]{Lemma}
\usepackage{empheq} 
\usepackage{mathtools}
\usepackage{xcolor}
\usepackage{mathrsfs}
\usepackage{amsfonts}
\usepackage[unicode=true,
 bookmarks=false,
 breaklinks=false,pdfborder={0 0 1},backref=section,colorlinks=false]
 {hyperref}

\makeatletter
\usepackage{graphicx}
\usepackage{fancybox}
\usepackage{comment}
\usepackage{float}
\usepackage{cite}
\usepackage{enumitem}
\usepackage{bbold}
\usepackage{multicol}
\usepackage{subfigure}
\usepackage{algorithm,algcompatible,amsmath}
\algnewcommand\INPUT{\item[\textbf{Input:}]}%
\algnewcommand\VOID{\item[\textbf{            }]}%
\algnewcommand\OUTPUT{\item[\textbf{Output:}]}%
\algnewcommand\START{\item[\textbf{Start:}]}%
\algnewcommand\RETURN{\item[\textbf{Return:}]}%

\definecolor{TB}{rgb}{0.6,0,0}
\definecolor{SG}{rgb}{0.6,0,0.4}
\definecolor{LD}{rgb}{0,0.4,0.0}
\newcommand{\sgt}[1]{{\color{SG}(SGT: #1)}}
\newcommand{\luca}[1]{{\color{LD}(LUCA: #1)}}

\newcommand{\data}[1]{\breve{#1}}
\newcommand{\junk}[1] {}
\makeatother

\newcommand{\vet}[1]{\boldsymbol{#1}} 
\newcommand{\mat}[1]{\mathbf{#1}} 
\newcommand{\mati}[1]{\boldsymbol{#1}} 

\newcommand{\eye}{\mat{I}}

\renewcommand{\Re}[1]{\mathrm{Re}\left \{#1\right\} }

\newcommand{\Real}{\mathbb{R}}
\newcommand{\Complex}{\mathbb{C}}

\newcommand{\smallsignal}[1]{\widetilde{\vet{#1}}}
\newcommand{\smallsignalder}[1]{\dot{\widetilde{\vet{#1}}}}
\newcommand{\smallsignalmat}[1]{\widetilde{\vet{#1}}}
\newcommand{\bias}[1]{{\vet{#1}_{\!0}}}

\newcommand{\RTVF}{{\tt RTVF}}
\newcommand{\TDVF}{{\tt TDVF}}

\newcommand{\heaviside}{\Theta}

\begin{document}

\title{Handling Initial Conditions in Vector Fitting for Real Time Modeling of Power System Dynamics}

\author{Tommaso~Bradde,~\IEEEmembership{Student Member,~IEEE},
    Samuel Chevalier,~\IEEEmembership{Student Member,~IEEE}
    Marco~De~Stefano,~\IEEEmembership{Student Member,~IEEE},
    Stefano Grivet-Talocia,~\IEEEmembership{Fellow,~IEEE},
	Luca Daniel,~\IEEEmembership{Senior Member,~IEEE}%
	}

\maketitle

\begin{abstract}
This paper develops a predictive modeling algorithm, denoted as Real-Time Vector Fitting (\RTVF), which is capable of approximating the real-time linearized dynamics of multi-input multi-output (MIMO) dynamical systems via rational transfer function matrices. Based on a generalization of the well-known Time-Domain Vector Fitting (\TDVF) algorithm, \RTVF\ is suitable for online modeling of dynamical systems which experience both initial-state decay contributions in the measured output signals and concurrently active input signals. These adaptations were specifically contrived to meet the needs currently present in the electrical power systems community, where real-time modeling of low frequency power system dynamics is becoming an increasingly coveted tool by power system operators. After introducing and validating the \RTVF\ scheme on synthetic test cases, this paper presents a series of numerical tests on high-order closed-loop generator systems in the IEEE 39-bus test system.
\end{abstract}

\section{Introduction}
\label{sec:Introduction}
\IEEEPARstart{D}{ue} to the aggressive deployment of Wide Area Monitoring Systems (WAMS), a deluge of time series data streams are emerging from modernizing smart grids. For instance, as of 2017, Operating Procedure No. 22 specifies that all generation above 100MW in the ISO New England system must provide Phasor Measuring Unit (PMU) observability at the point of interconnection~\cite{Vaiman:2018}. Similarly, the 2018 Nodal Operating Guide specifies that all new generators above 20MVA in the ERCOT (Texas) system must provide PMU observability~\cite{ERCOT:2018}. In order to both capitalize on these data streams and enforce effective dynamic security assessment (DSA) in the face of a rapidly modernizing energy landscape with shrinking inertia margins~\cite{Milano:2018}, the North American Electric Reliability Corporation (NERC) has recently implemented new directives~\cite{NERC_Slides:2016} which mandate the continual development of static and dynamic network planning models. The simulated response of these models must be compared to actual time series data collected in the network to validate their accuracy~\cite{Akhlaghi:2020}. Associated procedures must also resolve model prediction aberrations. Traditional ``staged" testing of grid models can be costly and inconvenient, because generators must go offline~\cite{Li:2017}. Thus, performing model tuning and validation online is particularly valuable. 

Furthermore, real-time modeling of the grid's dynamics at a ``wide area" level can provide useful operational benefits, such as modal damping ratio monitoring and controller tuning. For instance, the 3.1 GW Pacific DC Intertie (PDCI) line in the Western United States has been recently outfitted with a damping controller to increase its transfer capabilities~\cite{Pierre:2019}. This celebrated controller is tuned using an open loop transfer function model of the power system which was empirically measured via intrusive multi-sine modulation of the HVDC power controllers~\cite{Trudnowski:2014}. Methods for online modeling of these types of dynamics, therefore, could allow for enhanced controller tuning across many wide area applications.

In the electromechanical frequency range, power system dynamics are still dominated by synchronous generators~\cite{Milano:2018}. A wealth of literature exists on the nonlinear physical modeling of these generators, their associated high-order controllers, and the networks to which they interconnect~\cite{Kundur:1994,Bialek:2020}. Applying, tuning, and validating these models in the context of large realistic power systems faces a variety of practical obstacles; these include the exclusion of unmodeled dynamics, measurement noise, uncertain physical parameter values, unknown controller changes, and, in the limiting case of linearized modeling, residual nonlinearities and drifting equilibrium points. In recent years, a variety of methods have been proposed for the purpose of identifying, validating and tuning these models~\cite{NERC_Slides:2016,Foroutan:2019,Huang:2018,Li:2017,Huang:2013,Wu:2018}, usually via Kalman filter-based parameter tuning and ``play-back" simulation. The vast majority of the proposed modeling and validation algorithms are characterized by two salient features. First, they are parameterized by a physical prior; that is, they leverage the structure of a given physical model which can assumably be tuned to explain the full set observed dynamics. Second, they are typically designed to be deployed in the presence of a sufficiently strong network perturbation (e.g., ground fault). Together, these assumptions allow for a variety of helpful simplifications, such as initial state decay negligence and high-fidelity prior availability. Such algorithmic features can be burdensome if the prior model is incomplete or if real-time modeling updates are desired.

Predictive modeling algorithms which are free from these limitations can provide a variety of advantages. If power system modeling can be performed by a parameterized ``black box" approximation, then the associated algorithm will be freed from the constraints of a potentially erroneous, or completely unknown, prior model. Additionally, if this algorithm can be implemented in the face of ambient, rather than severe, operating conditions, then updates can be performed on the fly. There is a large collection of well-cited literature devoted to the creation of linear dynamical black box models from time series data in power systems, primarily at the ``wide area" level, but also at the individual generator level. Excellent reviews on the topic are provided in~\cite{ Pierre:2012, Dosiek:2013 }. One of the most popular ``modal identification" methods, known as Prony analysis, was first applied to power systems in 1990~\cite{ Hauer:1990}. It was further shown in~\cite{ Pierre:1997} that Prony analysis applied to transient data and Wiener-Hopf linear prediction applied to ambient data could both identify very similar system dynamics. Subsequent publications further refined these methods~\cite{Wies:2003,DosiekEst:2013}, but most of them were focused on identification of system modes (i.e. poles), rather than the full transfer functions. More recent work has focused on building model reduced transfer functions from measured terminal data. Using a simple autoregessive with exogenous input (ARX) model, \cite{Chai:2016} develops a process for computing a dynamic equivalent transfer function with simulated or measured data. Voltage magnitude and frequency are chosen as inputs, whiles active and reactive power flows across lines are chosen as outputs. MIMO transfer function approaches are developed in~\cite{Zhang:2017}\junk{Zhang:2018}, where poles, zeros, and model order are iteratively perturbed based on a least-squares numerical procedure inside a standard MATLAB toolbox.

Most recently, vector fitting approaches have been leveraged to construct power system models from measured field data. In~\cite{ Banuelos-Cabral:2019 }, authors use the numerical Laplace transform to push the problem into the frequency domain. Next, vector fitting and relative dominant-pole measurements (RDPMs) are iteratively leveraged to build estimations of wide area system modes. Similarly,~\cite{Schumacher:2018} combines time domain vector fitting and the popular ring down analysis to develop a novel modal estimation routine. In a more generic application, \cite{Jakobsen:2017} uses out-of-the-box \TDVF\ to construct a SISO linear model of a turbine governor system from measured PMU; droop gain values are then captured from the constructed model.

The black-box identification of dynamic models from time series data goes far beyond power system applications and is regarded as a well established field~\cite{mon-2015-PM,candy2019model}; {popular algorithms include the Multivariable Output Error State Space (MOESP) and Numerical algorithms for Subspace State Space System Identification (N4SID) techniques~\cite{Jamaludin:2013}.} As noted, one prominent approach is the Time-Domain Vector Fitting (\TDVF) scheme~\cite{jnl-2003-mwcl-TDVF,jnl-2004-aeu-TDVF,mon-2015-PM}. This approach is applicable for estimating rational models of linear systems, starting from the time domain samples of input and output data. Unfortunately, canonical \TDVF\ has the following restrictions:
\begin{itemize}
    \item The system must be at rest when the data acquisition begins. This guarantees that the input and output data are related by a linear transfer function with no contribution from the zero-input (i.e. initial state decay) response.
    \item For a MIMO system with $P$ ports, only separate modeling of each column of the transfer matrix is possible. This limitation requires that only one of the $P$ input components is acting on the system during the separate modeling periods for each transfer matrix column.
\end{itemize}
Since online modeling typically requires the unknown system to be characterized (i) during its dynamic evolution and (ii) in the presence of concurrently acting inputs, the above limitations often make \TDVF\ unsuitable for real-time MIMO applications. {This is especially true in the case of electrical power systems. Due to constant load switching and system perturbations, the system never reaches a truly steady state condition. Thus, any measured signals from the system will always contain some degree of initial state decay. Furthermore, generic subsystems of three-phase AC power systems are inherently MIMO\footnote{{This is still true, even when a symmetric three-phase power system is collapsed down to a so-called single-line diagram. Since single-line diagrams are modeled using a quasi-stationary (e.g., sinusoidal) phasor approximation~\cite{Kundur:1994}, there are always two components associated with the input and output signals at the terminal of any power system component. In this quasi-stationary context, only under very special symmetry conditions can a MIMO transfer function be fully represented as a SISO transfer function~\cite{Harnefors:2007}.}} in nature~\cite{Harnefors:2007}, with orthogonal input signals (e.g., voltage magnitude and voltage phase) which are always applied in a concurrent manner.}

{In order to address the aforementioned limitations of canonical \TDVF}, this paper introduces a generalized vector fitting extension, known as Real-Time Vector Fitting (\RTVF). The remainder of this paper is organized as follows. In Section~\ref{sec:Background}, we recall the basic \TDVF\ scheme in order to provide a framework on which the proposed approach is built. We then present our key contributions, namely: 
\begin{enumerate}
\item We remove the restrictions of the basic \TDVF\ scheme by proposing a new problem setting {(Section~\ref{sec:NewSetting})}.
\item We develop a first generalization of the basic \TDVF\ scheme by removing the requirement of pure zero-state conditions, hence allowing for the presence of non-vanishing initial conditions {(Section~\ref{sec:free evolution})}.
\item We develop a second generalization by allowing all input components to act concurrently in the time series data sets used to train the \RTVF\ model {(Section~\ref{sec:RTVF})}.
\end{enumerate}
In Section~\ref{sec:Implementation}, we provide implementation details\junk{\color{SG}, including a discussion on computational cost}. Numerical test results, based on both synthetic and realistic power system models, are given in Section~\ref{sec:Results}.


\section{Technical Background}
\label{sec:Background}

\subsection{Notation}

We denote a generic scalar as $x$, a generic vector as $\vet{x}$ or $\bias{X}$, and a generic matrix as $\mat{X}$.  The identity matrix is denoted as $\mat{I}$, with size inferred from the context. We use the symbol $s$ for the complex frequency (Laplace) variable; $\Real$ and $\Complex$ represent the sets of real and complex numbers, respectively.

We consider a possibly nonlinear dynamic system $\mathcal{S}$ with input and output signals denoted as $\vet{u}(t)\in\Real^P$ and $\vet{y}(t)\in\Real^P$, respectively. For later use, we denote with $\vet{x}(t)\in \Real^{ \bar{N}}$ some unknown system state vector, although we assume no information on the internal system representation. We assume instead that a measurement tool is available that returns $K$ samples of time-domain input-output vectors
\begin{equation}\label{eq:data}
    \vet{u}(t_k), \vet{y}(t_k) \quad k=1,...,K
\end{equation}
acquired at sampling rate $F_s$. Without loss of generality, we  set $t_1=0$ throughout this paper. All derivations will hold true for $t_1\neq 0$, provided the time variable is redefined as $t\leftarrow t-t_1$.

\subsection{The Standard Setting for Data-Driven Modeling}
\label{sec:StandardSetting}
When the underlying system $\mathcal{S}$ is Linear and Time-Invariant (LTI), Laplace-domain input and output signals are related by
\begin{equation}\label{eq:ssTransferFunction}
\vet{Y}(s) = \data{\mat{H}}(s) \vet{U}(s).
\end{equation}
An estimate $\mat{H}(s)$ of the true transfer function $\data{\mat{H}}(s)$ can be determined from the samples~\eqref{eq:data} through one of the several available data-driven model order reduction methods.\junk{When the underlying system $\mathcal{S}$ is Linear and Time-Invariant (LTI), an estimate $\mat{H}(s)$ of the transfer function $\data{\mat{H}}(s)$ relating the Laplace-domain input and output signals through
\begin{equation}\label{eq:ssTransferFunction}
\vet{Y}(s) = \data{\mat{H}}(s) \vet{U}(s)
\end{equation}
can be determined from the  samples~\eqref{eq:data} through one of the several available data-driven model order reduction methods.}
In particular, the \TDVF\ scheme~\cite{mon-2015-PM,jnl-2003-mwcl-TDVF,jnl-2004-aeu-TDVF} considered in this work assumes that the system is initially at rest, and the \emph{initial conditions} vanish identically as
\begin{equation}\label{eq:init_cond_0}
    \vet{u}(0)\equiv \vet{0}, \quad \vet{y}(0)\equiv \vet{0}, \quad \vet{x}(0) \equiv \vet{0}.
\end{equation}
This setting guarantees that only the zero-state response contribution is present in the output samples.

\subsection{Time-Domain Vector Fitting}\label{sec:tdvf}

The basic \TDVF\ scheme assumes availability of
\begin{itemize}
    \item time series of each output $y_{ij}(t)$ at port $i$ excited by a single input $u_j(t)$ placed at port $j$ and acting alone, with $u_{k\neq j}=0$; this requirement imposes a restriction on the training sequences that can be used for model extraction;
    \item some initial estimate of the dominant system poles $\{q_n,\, n=1,\dots ,N\}$. Usually, such poles are initialized as random real or complex conjugate pairs with $\Re{q_n}<0$ and $|q_n|<\Omega$, where $\Omega$ is the modeling bandwidth of interest~\cite{mon-2015-PM,tpwrd-1999-VF}.
\end{itemize}
Based on the training data, \TDVF\ constructs the approximation
\begin{equation}\label{eq:tdvf_ij}
    d_0 \cdot {y}_{ij} (t) + \sum^N_{n=1} d_n \cdot {y}_{ij}^{(n)} (t) \approx c_{ij}^{(0)} \cdot {u}_j(t) + \sum^N_{n=1}   c^{(n)}_{ij} \cdot  {u}^{(n)}_j(t)
\end{equation}
for $t=t_k$ with $k=1,\dots,K$, where $c_{ij}^{(n)}$ and $d_n$ are unknown coefficients to be determined via a linear least squares solution. In~\eqref{eq:tdvf_ij}, superscript ${}^{(n)}$ denotes the result obtained by the single-pole filter (i.e. convolution) via
\begin{equation}
    z^{(n)}(t) = \int_{0}^t e^{q_n(t-\tau)} z(\tau) d \tau \label{eq:filtered_z}
\end{equation}
on any arbitrary signal $z(t)$ with $z(0)=0$. The regression problem~\eqref{eq:tdvf_ij} corresponds to the frequency-domain relation
\begin{equation}\label{eq:siso condition}
    {Y}_{ij}(s) \approx H_{ij}(s)\, {U}_j(s)  \approx \frac{c^{(0)}_{ij}+\sum^N_{n=1}\dfrac{c^{(n)}_{ij}}{s-q_n}}{d_0+\sum^N_{n=1}\dfrac{d_n}{s-q_n} } \,\cdot\, {U}_j(s) 
\end{equation}
which provides an element-wise rational approximation $H_{ij}(s)\approx \data{H}_{ij}(s)$ written in barycentric form. The initial poles $q_n$ in~\eqref{eq:siso condition} cancel out, and the actual poles of $H_{ij}(s)$ correspond to the zeros $z_n$ of the denominator
\begin{equation}\label{eq:denominator}
    D(s)=d_0+\sum^N_{n=1}\dfrac{d_n}{s-q_n}.
\end{equation}
Problem~\eqref{eq:tdvf_ij} is solved iteratively, by using these zeros as starting poles for the next iteration via $q_n \leftarrow z_n$. Iterations stop when the poles and/or the least-squares residual (fitting error) stabilize~\cite{mon-2015-PM,6470726}, or alternatively when a maximum number of iterations $\nu_{\max}$ is reached. 
\junk{
A good proxy for pole convergence is provided by the norm of the vector $\vet{d}' = [d_1, \dots, d_N]$ collecting all denominator coefficients excluding $d_0$. When such coefficients approach zero, the denominator function $D(s)$ approaches a constant value, implying that the poles and the zeros of $D(s)$ are nearly coincident. Under this condition, the poles become invariant through iterations and convergence is attained. Therefore, the iterations can be stopped when
\begin{equation}\label{eq:denCoeffStop}
\sqrt{ \textstyle\sum_{i=1}^N d_i^2 } \leq  \epsilon
\end{equation}
where $\epsilon$ is a desired tolerance. Alternatively, a maximum number of algorithm iterations $\nu_{\max}$ can be set.}
\section{A New Problem Setting for TDVF}
\label{sec:NewSetting}

Let us now consider system $\mathcal{S}$ operating in real time, with input and output measured signals~\eqref{eq:data} collected during system operation. Since data recording may start at an arbitrary time instant when system is not at rest, the output samples $\vet{y}(t)$ may include contributions from both the zero-input and the zero-state response, implying that all initial conditions cannot be considered as vanishing as in~\eqref{eq:init_cond_0}. Moreover, all input channels are expected to be active concurrently. Therefore, the basic \TDVF\ assumptions do not hold and need a generalization.

From now on, we assume a mildly nonlinear systems $\mathcal{S}$, whose dynamics can be approximated~\cite{jnl-2020-tcas-bid} by splitting inputs, outputs and states as
\begin{equation}\label{eq: small signals}
\begin{aligned}
    \vet{u}(t) & = \bias{U} + \smallsignal{u}(t), \\
    \vet{y}(t) & = \bias{Y} + \smallsignal{y}(t), \\
    \vet{x}(t) & = \bias{X} + \smallsignal{x}(t),
\end{aligned}
\end{equation}
where
\begin{equation}\label{eq:init_cond}
    \bias{U}=\vet{u}(0), \quad \bias{X}=\vet{x}(0), \quad \bias{Y}=\vet{y}(0)
\end{equation}
are regarded as \emph{non-necessarily} vanishing initial conditions. The evolution of the \emph{small-signal} components $\smallsignal{u}(t)$, $\smallsignal{y}(t)$, $\smallsignal{x}(t)$ can be accurately related by an LTI operator with transfer function $\data{\mat{H}}(s)$. Our main objective is to devise a numerical scheme that, based on the samples~\eqref{eq:data}, returns an estimate $\mat{H}(s)\approx \data{\mat{H}}(s)$ of the small-signal transfer function. This objective is attained by exploiting two generalizations of the basic \TDVF\ scheme:
\begin{enumerate}
    \item We remove the requirements of pure zero-state conditions by allowing for the presence of nonvanishing initial conditions~\eqref{eq:init_cond}. This problem is addressed in Sec.~\ref{sec:free evolution}.
    \item We allow for all input components ${u}_j$ acting concurrently in the training time series, as in common system operation conditions. This problem is addressed in Sec.~\ref{sec:RTVF}.
\end{enumerate}


\section{Handling Initial conditions}
\label{sec:free evolution}

 In this section, 
 we formulate a generalized identification problem that provides an estimate $\mat{H}(s)$ of the small-signal transfer function, assuming non-vanishing initial conditions. \junk{We remark that,} Although initial conditions $\bias{U}$, $\bias{Y}$ are known from the training data, no information on the initial state $\bias{X}$ is available. Thus, a direct decomposition of $\vet{y}(t)$ into its zero-state and zero-input contributions is generally not possible. This section provides theoretical justification for the proposed \RTVF\ formulation of Section~\ref{sec:RTVF} and may be skipped at first reading.

In order to characterize the role of the unknown initial state, we first consider a generic LTI system in state-space form:
\begin{equation}\label{eq:ss}
\begin{aligned}
    \dot{\vet{x}}(t) & = \mat{A}\vet{x}(t) + \mat{B}\vet{u}(t), \\
    \vet{y}(t) & = \mat{C}\vet{x}(t) + \mat{D}\vet{u}(t),
\end{aligned}
\end{equation}
with the only restriction that $\mat{A}$ should be nonsingular. Inserting the signal decomposition~\eqref{eq: small signals} into~\eqref{eq:ss} leads to
\begin{align}
        \dot{\vet{x}}(t) = \smallsignalder{x}(t) & = \mat{A}(\bias{X}+\smallsignal{x}(t)) + \mat{B}(\bias{U}+\smallsignal{u}(t))     \label{eq:ssDecomposition} \\
        \smallsignal{y}(t) + \bias{Y} & = \mat{C}(\bias{X}+\smallsignal{x}(t)) + \mat{D}(\bias{U}+\smallsignal{u}(t)).
        \label{eq:SSDecomposition_output}
\end{align}
The output~\eqref{eq:SSDecomposition_output} can be equivalently rewritten by splitting the constant and the time-varying small-signal components as
\begin{align}
    \bias{Y} & = \mat{C}\bias{X} + \mat{D}\bias{U}, \label{eq:outputDecomposition_0}\\
    \smallsignal{y}(t) & = \mat{C}\smallsignal{x}(t) + \mat{D}\smallsignal{u}(t), \quad \forall t \geq 0. \label{eq:outputDecomposition_t}
\end{align}
Two scenarios are possible:
\begin{enumerate}
    \item The system is at constant steady-state\footnote{This scenario is common in electronic circuit simulation, where a constant bias is applied first and all initial conditions are found; transient analysis is performed next, starting from the computed initial conditions. {Since electrical power systems never reach steady state, this condition can never be exploited in practice.}} for $t=0$ (equivalently, $\forall t\leq 0$). Under this assumption, all small-signal components vanish\junk{identically} for $t\leq 0$. Therefore, \eqref{eq:ssDecomposition} reduces to
    \begin{equation}\label{eq:DC_state}
        \mat{A} \bias{X} + \mat{B} \bias{U} = \vet{0} \quad \rightarrow \quad \bias{X} = - \mat{A}^{-1} \mat{B} \bias{U}
    \end{equation}
    and provides the initial state condition $\bias{X}$. Since $\mat{A}$ is nonsingular, the system has no poles at the origin and supports constant steady-state operation. Combining~\eqref{eq:ssDecomposition} with~\eqref{eq:DC_state}, for $t>0$, the small-signal components fulfill the standard dynamic equation
    \begin{equation}\label{eq:ss_small_signal}
        \smallsignalder{x}(t) = \mat{A} \smallsignal{x}(t) + \mat{B} \smallsignal{u}(t).
    \end{equation}
    Combining \eqref{eq:outputDecomposition_t} and \eqref{eq:ss_small_signal} provides the small-signal transfer function $\data{\mat{H}}(s)$ in terms of the state-space matrices:
    \begin{equation}\label{eq:tf_data}
        \data{\mat{H}}(s) = \mat{C}(s\eye - \mat{A})^{-1}\mat{B} + \mat{D} = \frac{\data{\mat{N}}(s)}{\data{D}(s)}
    \end{equation}
    with $\data{D}(s)=|s\mat{I}-\mat{A}|$.
    \junk{Since $\vet{y}(t) = \bias{Y} + \smallsignal{y}(t)$,}Identification of a rational model for $\data{\mat{H}}(s)$ can be performed by subtracting the initial conditions $\bias{U}$, $\bias{Y}$ from the input and output signals and then applying a zero-state identification scheme, such as basic \TDVF, to small-signal components $\smallsignal{u}(t)$, $\smallsignal{y}(t)$.
    \item The second scenario is relevant for our application, and it corresponds to the case where the system is \emph{not} operating under constant steady-state conditions for $t<0$. In this setting,~\eqref{eq:DC_state} does not hold and
    \begin{equation}
        \bias{X} \neq - \mat{A}^{-1} \mat{B} \bias{U}.
    \end{equation}
    Therefore, even if the initial conditions $\bias{U}$, $\bias{Y}$ are removed from the input and output signals, the corresponding small-signal output $\smallsignal{y}(t)$ still includes a contribution from the initial state. This contribution is analyzed next.
\end{enumerate}

\subsection{Characterization of residual zero-input contributions}
\label{sec:non_steaty_state}

Assuming that $\bias{U}$ and $\bias{X}$ are known, system evolution in terms of small-signal state components is obtained by integrating the dynamic equation~\eqref{eq:ssDecomposition} for $t>0$ as
\begin{equation}\label{eq: small signal state}
    \smallsignal{x}(t) = \int^t_{0} e^{\mat{A}(t-\tau)}(\mat{B}\smallsignal{u}(\tau) +  \mat{A}\bias{X} + \mat{B}\bias{U}) d\tau.
\end{equation}
Simple algebraic manipulations allow us to rewrite~\eqref{eq: small signal state} as
\begin{equation}\label{eq: state time solution}
\begin{aligned}
    \smallsignal{x}(t)= & \int^t_{0} e^{\mat{A}(t-\tau)}\mat{B}\smallsignal{u}(\tau)d\tau \\
    & + \left[e^{\mat{A}t} - \eye\right] \underbrace{(\bias{X} + \mat{A}^{-1}\mat{B}\bias{U})}_{\bias{T}},
\end{aligned}
\end{equation}
where the contribution of the initial state condition is explicit. Note that, in case of steady-state operation for $t<0$, the second term in~\eqref{eq: state time solution} vanishes since $\bias{T}=\vet{0}$, and the corresponding solution reduces to the solution of the small-signal system~\eqref{eq:ss_small_signal}. The term $\bias{T}$ can thus be considered as the difference between the actual initial state $\bias{X}$ and the constant state that would be obtained if the system were operating under steady-state conditions excited by the constant input $\bias{U}$.

Taking the Laplace transform of~\eqref{eq: state time solution} yields
\begin{equation}\label{eq:Yt_laplace}
    \smallsignalmat{X}(s) = (s\eye - \mat{A})^{-1}\mat{B}\,\smallsignalmat{U}(s) + \left[(s\eye - \mat{A})^{-1} -s^{-1} \eye \right]\bias{T}.
\end{equation}
Inserting~\eqref{eq:Yt_laplace} into the output equation~\eqref{eq:outputDecomposition_t} leads to
\begin{equation}\label{eq:laplace small signal}
    \smallsignalmat{Y}(s) = \mat{C}\smallsignalmat{X}(s) + \mat{D}\smallsignalmat{U}(s) = 
        \mat{\data{H}}(s)\smallsignalmat{U}(s) +  \mati{\Gamma}_0(s),
\end{equation}
where $\data{\mat{H}}(s)$ is given by~\eqref{eq:tf_data}. Additionally,
\begin{equation}
    \mat{\Gamma}_0(s) = (\mat{C}(s\eye - \mat{A})^{-1} -s^{-1} \mat{C})\,\bias{T} = \frac{\data{\vet{G}}(s)}{s\cdot\data{D}(s)},
\end{equation}
where $\data{\vet{G}}(s)$ is an unknown polynomial vector. Relation~\eqref{eq:laplace small signal} is therefore equivalent to
\begin{equation}\label{eq:nh_output_Laplace}
    \smallsignalmat{Y}(s) =\frac{\data{\mat{N}}(s)}{\data{D}(s)} \,\smallsignalmat{U}(s) + \frac{\data{\vet{G}}(s)}{s\cdot \data{D}(s)}\,.
\end{equation}
The two terms in~\eqref{eq:nh_output_Laplace} share the same denominator $\data{D}(s)$ up to a pole at $s=0$, which represents the constant contribution of the non-vanishing initial conditions. This observation is the key enabling factor for building a self-consistent vector fitting scheme to estimate model $\mat{H}(s)\approx \data{\mat{H}}(s)$, as it properly takes into account the presence of the additional term $\mat{\Gamma}_0(s)$ in~\eqref{eq:laplace small signal}.


\section{The Real-Time Vector Fitting scheme}
\label{sec:RTVF}

In this section, we propose our Real-Time Vector Fitting (\RTVF) scheme. In particular, our key idea and the proposed  generalizations follow:
\begin{enumerate}
    \item   {\bf The key idea} of our method consists of  
         adding to the standard rational transfer function expression an extra term that shares the same denominator $D(s)$ plus a pole at $s=0$, representing the constant contribution of the non-vanishing initial conditions.
    \begin{equation}\label{eq:RTVF fitting condition}
        \smallsignalmat{Y}(s) \approx \frac{{\mat{N}}(s)}{{D}(s)} \,\smallsignalmat{U}(s) + \frac{\vet{G}(s)}{s\cdot {D}(s)}\,.
    \end{equation}
   \item
    Since polynomials $\mat{N}(s)$ and $D(s)$ are typically expanded in a standard barycentric form as in~\eqref{eq:siso condition} using an initial pole set $q_n$, we propose the use of a similar expansion for the components of the unknown vector $\vet{G}(s)$:
    \begin{equation}\label{eq:VFfreeResp}
         G_i(s) = b^{(0)}_{i}+\sum^N_{n=1}\frac{b^{(n)}_{i}}{s-q_n} \quad \forall i = 1,\dots, P\,.
    \end{equation} 
    \item We further propose to account for multiple inputs potentially exciting the system simultaneously by applying linear superposition and expressing each output component $\widetilde{Y}_i(s)$ in terms of all input components $\widetilde{U}_j(s)$.
\end{enumerate}
The above considerations lead us to parameterize and formulate the fitting condition~\eqref{eq:RTVF fitting condition} as
\begin{align} \label{eq:RTVF_Laplace}
    \tilde{Y}_i(s) \approx& \sum^P_{j=1}\, \frac{c^{(0)}_{ij}+\sum^N_{n=1}\dfrac{c^{(n)}_{ij}}{s-q_n}}{d_0+\sum^N_{n=1}\dfrac{d_n}{s-q_n} }  \,\cdot\, \tilde{U}_j(s) \,+ \nonumber \\ &+\dfrac{b^{(0)}_{i}+\sum^N_{n=1}\dfrac{b^{(n)}_{i}}{s-q_n}}{s\cdot \left(d_0+\sum^N_{n=1}\dfrac{d_n}{s-q_n}\right)} \quad \forall i = 1,\dots, P.
\end{align}
In~\eqref{eq:RTVF_Laplace}, the coefficients $c_{ij}^{(n)}$ represent the elements in the numerator $\mat{N}(s)$ of the small-signal model transfer function $\mat{H}(s)$ expressed in barycentric form. The coefficients $d_n$ of the denominator are common to all transfer matrix entries, so that a common pole set is enforced for the model. Finally, the coefficients $b_i^{(n)}$ provide a parameterization of the zero-input response in barycentric form, as written in~\eqref{eq:RTVF fitting condition}.

Multiplying both sides by the common denominator and taking the inverse Laplace transform leads to the following time-domain fitting condition for $t\geq 0$:
\junk{\begin{align}\label{eq:TDVF linearization}
\begin{split}
    &d_0 \cdot \tilde{y}_i (t) + \sum^N_{n=1} d_n \cdot \tilde{y}_i^{(n)} (t) \\
    & \approx \sum^P_{j=1} \left[ c_{ij}^{(0)} \cdot \tilde{u}_j(t) + \sum^N_{n=1}   c^{(n)}_{ij} \cdot  \tilde{u}^{(n)}_j(t)             \right]\nonumber \\
    & + b_i^{(0)} \cdot \heaviside(t) + \sum^N_{n=1} b_i^{(n)} \heaviside^{(n)}(t) \quad \forall i = 1,\dots, P,
\end{split}
\end{align}}
\begin{align}\label{eq:TDVF linearization}
    d_0 \cdot \tilde{y}_i &(t) + \sum^N_{n=1} d_n \cdot \tilde{y}_i^{(n)} (t) \nonumber\\
    \approx & \sum^P_{j=1} \left[ c_{ij}^{(0)} \cdot \tilde{u}_j(t) + \sum^N_{n=1}   c^{(n)}_{ij} \cdot  \tilde{u}^{(n)}_j(t)             \right]\nonumber \\
    & + b_i^{(0)} \cdot \heaviside(t) + \sum^N_{n=1} b_i^{(n)} \heaviside^{(n)}(t), \; \forall i = 1,\dots, P,
\end{align}
where $\heaviside(t)$ is the Heaviside unit step function.

\junk{
, filtered input $\tilde{u}_j(t)$ and output $\tilde{y}_i^{(n)} (t)$ signals are obtained as in~\eqref{eq:filtered_z}, and
\begin{equation}\label{eq: time basis functions}
\begin{aligned}
     g^{(n)}(t) &= \int_{0}^t e^{q_n(t-\tau)} \heaviside(\tau) d \tau \,.
     \end{aligned}
\end{equation}
}

\junk{
Starting from the available data, we can compute the samples of the filtered signals~\eqref{eq: time basis functions} (which can be regarded as known basis functions) at the corresponding time instants, and we obtain the following time domain fitting conditions, which are valid $\forall i = 1,...,P$ and $\forall k=1,...,K$
\begin{align}\label{eq: discrete fitting}
\begin{split}
    &\sum^P_{j=1} \left( c_{ij}^{(0)} \cdot \tilde{u}_j(t_k) + \sum^N_{n=1}   c^{(n)}_{ij} \cdot  \tilde{u}^{(n)}_j(t_k)             \right)  + \\
    &+b_i^{(0)} \cdot \heaviside(t_k-t_1) + \sum^N_{n=1} b_i^{(n)} g^{(n)}(t_k) +\\
    &-\left(d_0 \cdot \tilde{y}_i (t_k) + \sum^N_{n=1} d_n \cdot y_i^{(n)} (t_k)\right) \approx 0.
\end{split}
\end{align}
}


\section{Implementation}
\label{sec:Implementation}

In this section, we provide a compact and efficient formulation of the least squares formulation at the heart of the \RTVF\ routine\junk{\color{SG}, as well as a computational cost analysis}.
%
%
Writing~\eqref{eq:TDVF linearization} for $t=t_k$, with $k=1,\dots,K$, leads to the following \RTVF\ condition in matrix form:
\begin{equation}\label{eq:fitting all times}
    \mati{\phi}_i \cdot \vet{d} + \sum_{j=1}^P \mati{\psi}_j \cdot \vet{c}_{ij} + \mati{\beta} \cdot \vet{b}_i \approx \vet{0} \quad \forall i=1,2,...,P \,,
\end{equation}
where the vectors collecting the unknown coefficients are
\begin{align}
{\vet d}=\left[\begin{array}{c}
d_{0}\\
\vdots\\
d_{N}
\end{array}\right],\,{\vet c}_{ij}=\left[\begin{array}{c}
c_{ij}^{(0)}\\
\vdots\\
c_{ij}^{(N)}
\end{array}\right],\,{\vet b}_{i}=\left[\begin{array}{c}
b_{i}^{(0)}\\
\vdots\\
b_{i}^{(N)}
\end{array}\right],
\end{align}
and where the regressor matrices collecting the filtered signal time samples (according to~\eqref{eq:filtered_z}) are defined as
\begin{align}
    \mati{\phi}_i =-&\begin{bmatrix}
                   \tilde{y}_i(t_1)&\tilde{y}^{(1)}_i(t_1)&\hdots&\tilde{y}^{(N)}_i(t_1)\\
                   \vdots&\vdots&\ddots&\vdots\\
                   \tilde{y}_i(t_k)&\tilde{y}^{(1)}_i(t_K)&\hdots&\tilde{y}^{(N)}_i(t_K)
                   \end{bmatrix}\\
   \mati{\psi}_j =\quad&\begin{bmatrix}
                  \tilde{u}_j(t_1)&\tilde{u}^{(1)}_j(t_1)&\hdots&\tilde{u}^{(N)}_j(t_1)\\
                  \vdots&\vdots&\ddots&\vdots\\
                  \tilde{u}_j(t_k)&\tilde{u}^{(1)}_j(t_K)&\hdots&\tilde{u}^{(N)}_j(t_K)
                  \end{bmatrix}\\
   \mati{\beta} =\quad &\begin{bmatrix}
                  1 & \heaviside^{(1)}(t_1) & \hdots & \heaviside^{(N)}(t_1)\\
                  \vdots & \vdots & \ddots & \vdots\\
                  1 & \heaviside^{(1)}(t_K) & \hdots & \heaviside^{(N)}(t_K)
                  \end{bmatrix} \,.
\end{align}
Further, by defining
\junk{\begin{equation}
    \mati{\Delta}=\begin{bmatrix} \mati{\psi}_1 & \hdots &\mati{\psi}_P & \mati{\beta} \end{bmatrix}, \qquad 
    \vet{a}_i = \begin{bmatrix} \vet{c}_{i1} \\
                                      \vdots \\
                                      \vet{c}_{iP}\\
                                      \vet{b}_i
                       \end{bmatrix}
\end{equation}}
\begin{equation}
    \mati{\Delta}=\begin{bmatrix} \mati{\psi}_1 & \!\!\hdots\!\! &\mati{\psi}_P & \mati{\beta} \end{bmatrix}, \;\; 
    \vet{a}_i = \begin{bmatrix} \vet{c}_{i1}^T & \!\!\hdots\!\! & \vet{c}_{iP}^T & \vet{b}_i^T\end{bmatrix}^T
\end{equation}
and collecting all components,~\eqref{eq:fitting all times} reveals the bordered-block-diagonal structure of the \RTVF\ least squares system:
\begin{equation}\label{eq: LSproblem}
    \begin{bmatrix}
     \mat{\Delta} &     \mat{0}      & \hdots &     \mat{0} & \mati{\phi}_1 \\
          \mat{0}      &\mat{\Delta}& \hdots &     \mat{0}  & \mati{\phi}_2 \\
          \vdots   &    \vdots  & \ddots &  \vdots &   \vdots   \\
          \mat{0}      &      \mat{0}     & \hdots &\mat{\Delta} & \mati{\phi}_P
    \end{bmatrix}
    \begin{bmatrix}
    \vet{a}_1 \\ \vet{a}_2 \\ \vdots \\ \vet{a}_P \\ \vet d
    \end{bmatrix} \approx \mat{0}.
\end{equation}
Standard techniques can be employed to avoid the all-zero trivial solution, as explained in~\cite{tpwrd-2000-RVF,mon-2015-PM}. 

As in standard \TDVF, once the set of unknown coefficients is found by solving~\eqref{eq: LSproblem}, the zeros $z_n$ of the denominator $D(s)$ are computed and used as initial poles for the next iteration. The process is repeated until convergence. Pseudocode for \RTVF\ is provided in Algorithm~\ref{al:pseudocode}. The final steps (lines~8--10) estimate the residues of a rational approximation based on the fixed poles obtained from the pole relocation process (lines 1--7). In line~10, the vector $\smallsignal{\vet{y}}_i$ collects all time samples of the $i$-th small-signal output component.

\begin{algorithm}
		\caption{The \RTVF\ algorithm}\label{al:pseudocode}
		{\small 
		\begin{algorithmic}[1]
			\INPUT Time samples  $\vet{u}(t_k) ,\, \vet{y}(t_k) $, sampling frequency $F_s$, starting poles $\{q_1,\dots,q_N\}$, maximum iteration number $\nu_{\rm max}$
			\OUTPUT Estimated transfer function $\mat{H}(s)$
			\STATE Compute $\smallsignal{u}(t_k)\leftarrow \vet{u}(t_k)-\vet{u}(t_1)$, $\smallsignal{y}(t_k)\leftarrow \vet{y}(t_k)-\vet{y}(t_1)$
			\FOR {$\nu=1,..,\nu_{\rm max}$ }
			\STATE Compute filtered signals $\tilde y_i^{(n)}(t_k)$, $\tilde u_i^{(n)}(t_k)$, $\heaviside^{(n)}(t_k)$
			\STATE Build and solve the least squares problem~\eqref{eq: LSproblem}
			\STATE Compute the zeros $z_n$ of denominator $D(s)$ in~\eqref{eq:denominator}
			\STATE Set $q_n \leftarrow z_n$
			\ENDFOR
			\STATE Set $D(s)=1$
			\STATE Compute filtered signals $\tilde u_i^{(n)}(t_k)$, $\heaviside^{(n)}(t_k)$
			\STATE Build matrix $\mati{\Delta}$ 
 			and solve $\mati{\Delta} \vet{a}_i \approx \smallsignal{\vet{y}}_i$ for $i=1,\dots,P$
 			\STATE \textbf{return:} $\mat{H}(s) = \mat{N}(s)$, where $N_{ij}(s)$ is numerator of~\eqref{eq:siso condition}
		\end{algorithmic}}
\end{algorithm} 
\section{Numerical Results}
\label{sec:Results}
In this section, we test \RTVF\ in three different settings. First, we analyze its performance in the context of high order and high dimensional synthetic test systems. Next, we test its ability to model individual generator dynamics in the 39-bus system. Finally, we test its ability to model aggregated ``wide area" dynamics in the 39-bus system.
\subsection{Consistency} \label{subsec:consistency}

We tested the \RTVF\ consistency by running a systematic experimental campaign over a set of synthetic randomly generated LTI reference systems, with the objective of checking whether \RTVF\ could provide accurate estimates of all system poles. All modeled systems shared the same dynamic order of $N=10$, but had different sizes, with $P$ ranging from~2 to~30.
The set of input-output data were generated as colored noise which showed a flat power spectrum up to angular frequency $\omega_{\rm max}$, where the fastest pole of the reference system appeared. The sampling frequency was fixed to $F_s=10\,\omega_{\rm max}/2\pi$ , and the total number of collected samples was $K=5000$ in all cases. The modeling window started at sample $k=250$.

\begin{figure}
    \centering
    \includegraphics[width=0.9\columnwidth]{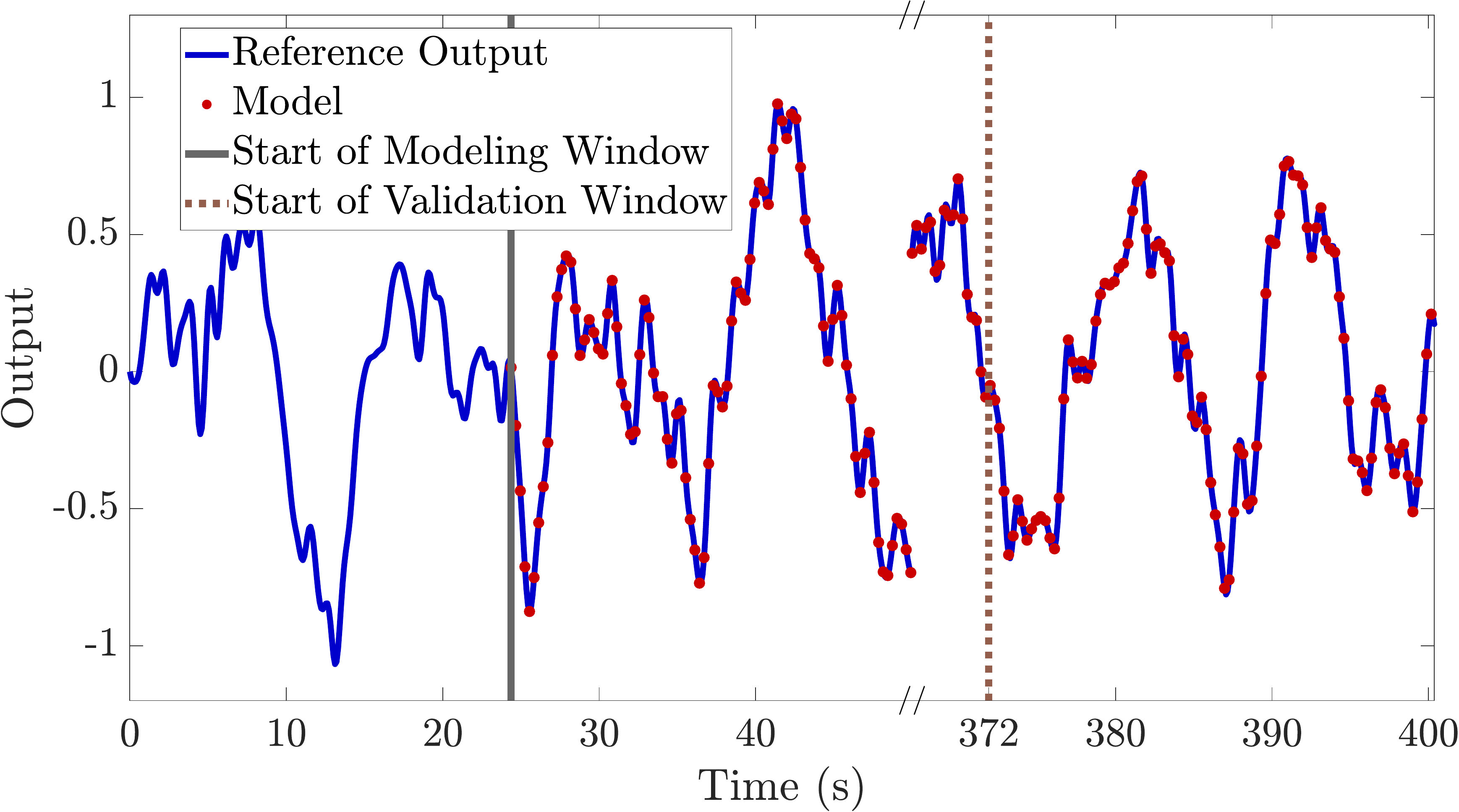}
     \caption{Time domain modeling results for a synthetic test case with $P=2$. The modeling window starts after 24~s, the validation window starts at 372~s.}
    \label{fig:inputOutputSignals}
\end{figure}
Three metrics were used to assess performance of \RTVF:
\begin{itemize}
    \item the consistency of the pole estimates, as measured by the Hausdorff distance\footnote{%
    The \emph{Hausdorff distance} between two sets $\mathcal{P}$ and $\mathcal{Q}$ is defined as
    \begin{equation}
    d_H(\mathcal{P},\mathcal{Q})=\max \{ \sup_{p \in \mathcal{P}} \, \inf_{q \in \mathcal{Q}} ||p-q|| , \,  \sup_{q \in \mathcal{Q}} \, \inf_{p \in \mathcal{P}} ||p-q|| \}.
    \end{equation}
    }\ %
    $d_H(\mathcal{P},\mathcal{Q})$ between the set of exact poles $\mathcal{P}= \{ p_1, \dots, p_{N} \}$ of the true system and the set of numerically computed poles $\mathcal{Q}= \{ q_1, \dots, q_{N} \}$;
    \item the worst-case time domain output error, computed as
    \begin{equation}
        E_{\infty}=\max_{i=1,\dots,P} ||\vet{y}_i-\data{\vet{y}}_i||_\infty,
    \end{equation}
    \item and the RMS-normalized maximum error, computed as
    \begin{equation}
        E^{\rm RMS}_\infty=\max_{i=1,\dots,P}  \frac{||\vet{y}_i-\data{\vet{y}}_i||_\infty}{||\data{\vet{y}}_i||_2}   .
    \end{equation} 
\end{itemize}

The experiments showed that \RTVF\ recovered the system poles almost exactly, with a set distance \mbox{$d_H(\mathcal{P},\mathcal{Q}) \leq 10^{-10}$} for all $29$ test cases. Similar results were obtained from the output errors: both $E_{\infty}$ and $E^{\rm RMS}_\infty$ were less than $10^{-11}$. As Fig.~\ref{fig:inputOutputSignals} shows, there is no practical difference between the model and the output data samples. In this idealized setting, we conclude that the performance of \RTVF\ is excellent across all investigated metrics.

To further test the consistency of \RTVF, we simulated the presence of measurement noise on the input and output signals (for the case where $P=2$, $N=10$). Signal corruption was performed by adding a vector of zero-mean Gaussian random variables $\vet{x}_n$ to any input or output small-signal vector  $\smallsignal{x}$ as
\begin{equation}
    \smallsignal{x}_N=\smallsignal{x}+\vet{x}_n,
\end{equation}
with a prescribed signal to noise ratio
\begin{equation}
\alpha={\rm SNR}=20\log\frac{{\rm RMS}\{\smallsignal{x}\}}{{\rm RMS}\{\vet{x}_n\}}.
\end {equation}

\begin{figure}
    \centering
    \includegraphics[width=0.9\columnwidth]{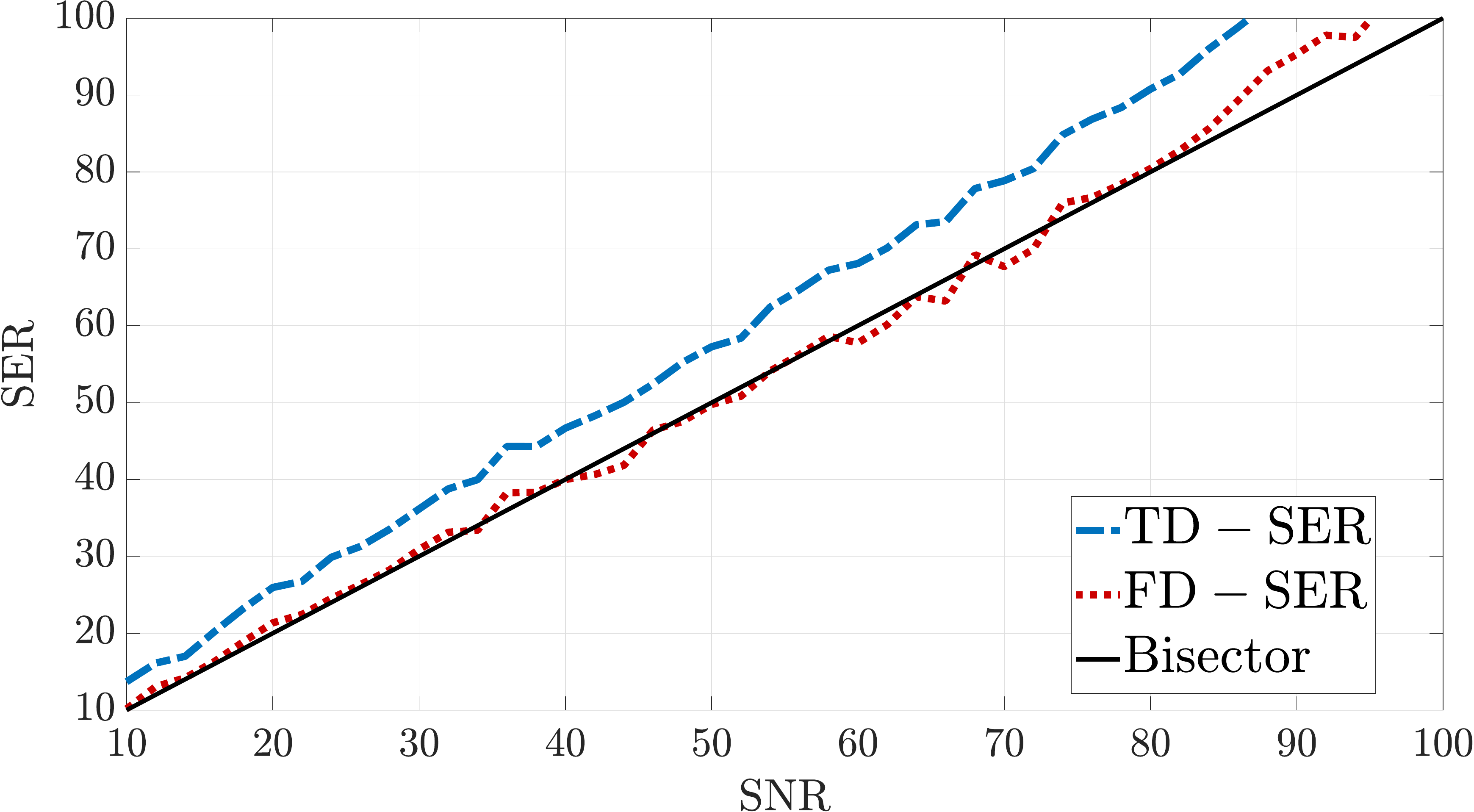}\\[4mm]
    \includegraphics[width=0.9\columnwidth]{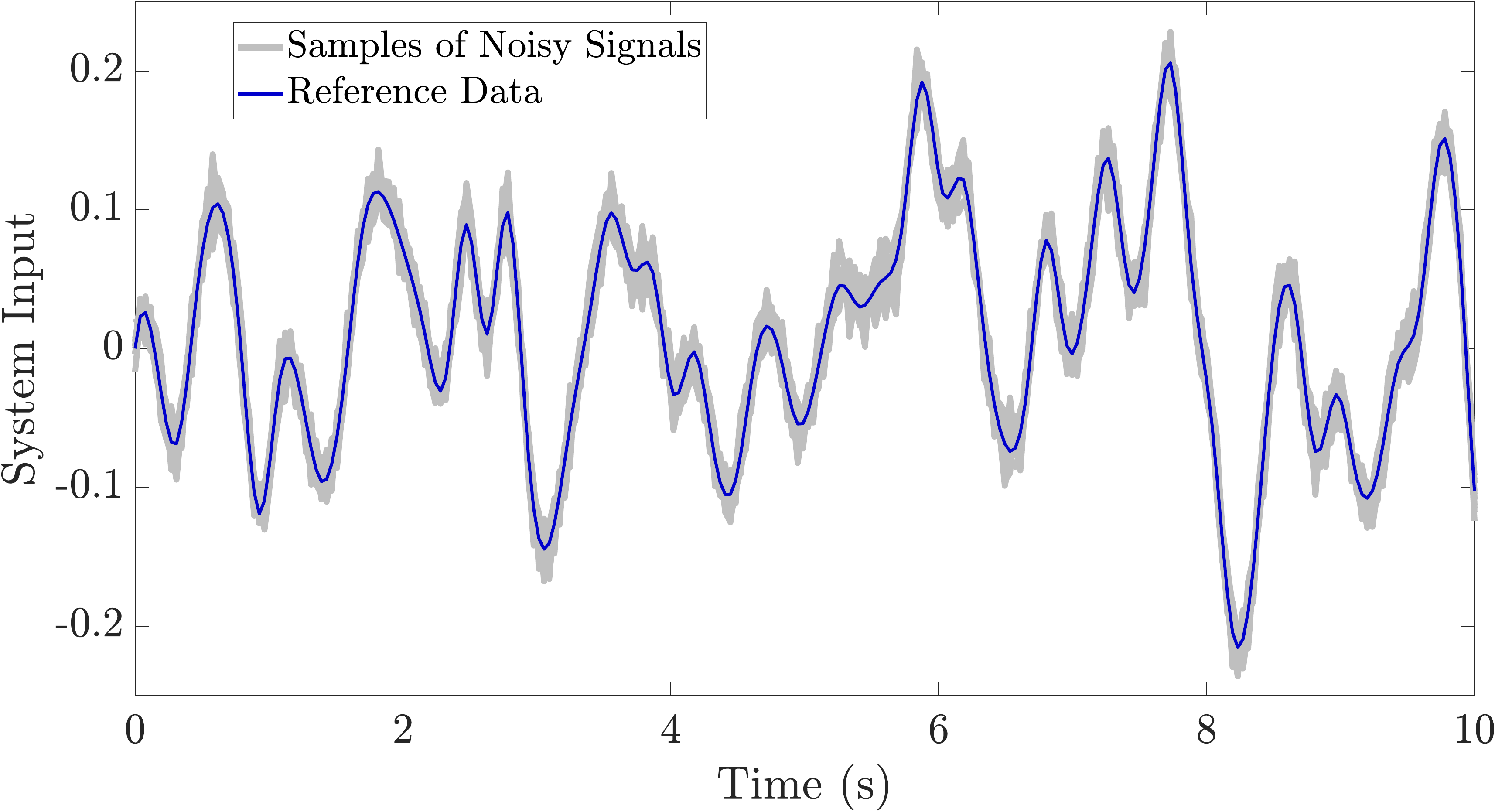}\\[3mm]
    \includegraphics[width=0.9\columnwidth]{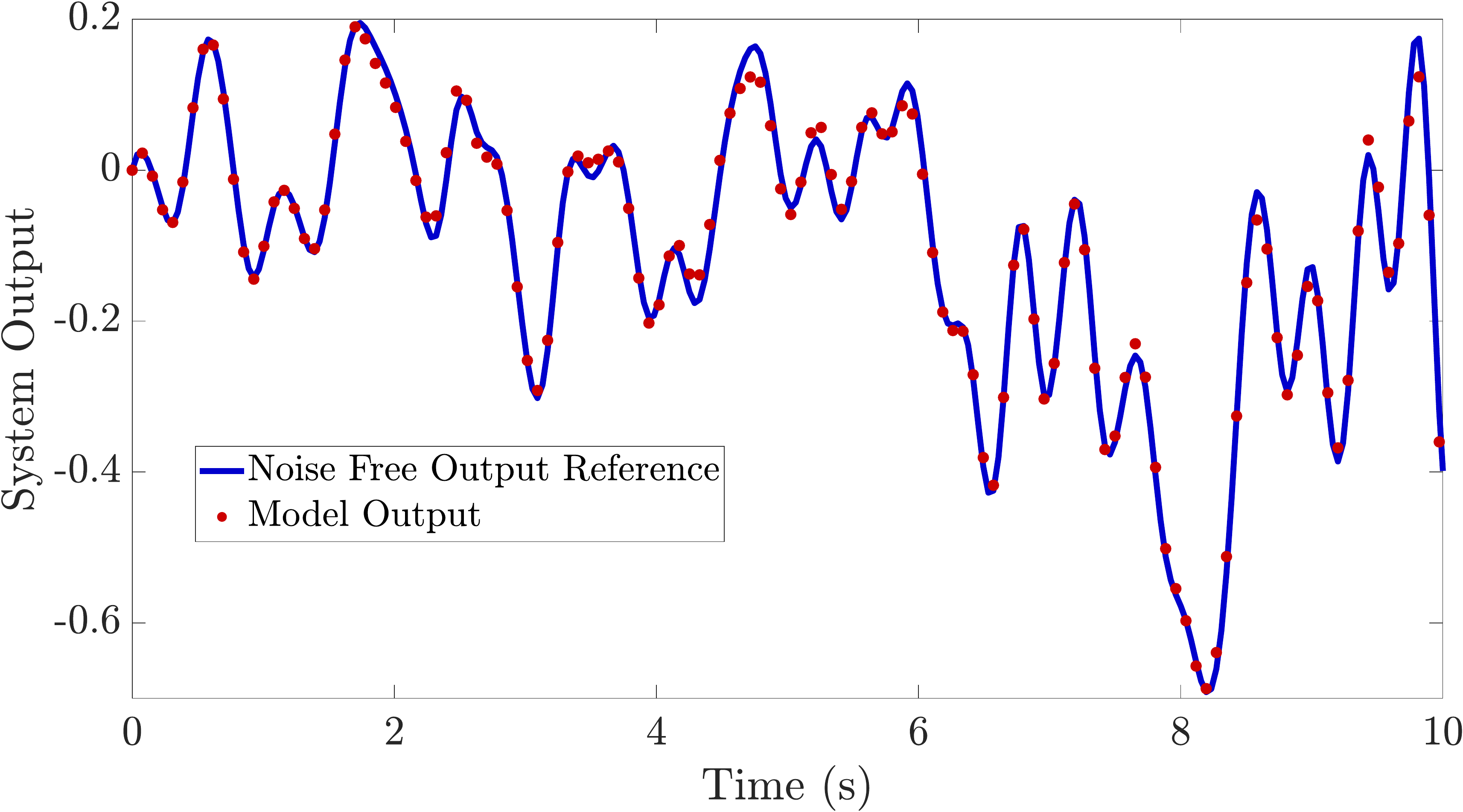}
     \caption{Top panel: the trend of $\rm{TD-SER}$ and $\rm{FD-SER}$ against the {\rm SNR}. Middle panel: corrupting one training input signal (solid line) with {\rm SNR=16} (a cloud of $R=50$ different realizations are depicted in a grey shade). Bottom panel: response of a time domain model extracted from one noisy data realization ({\rm SNR=16}) compared to the reference noise-free signal.}
     \label{fig:Ex_1_NoisyInputOutput}
\end{figure}

In our experiments, we considered increasing levels of ${\rm SNR}$, ranging from $10$ to $100$, with resolution steps of $2$. For each level of \rm{SNR}, we modeled $R=50$ different synthetic systems, and we computed the average \emph{Signal to Error Ratio} (SER) both in time and frequency domain, which is defined as follows. Let $\vet{z}$ be a vector collecting the samples of either a reference time-domain output signal or a target frequency-domain transfer matrix element, and $\vet{z}_M$ the corresponding response of one of the $R$ models. Then for this signal the SER is defined as
\begin{equation}\label{eq:SER}
     {\rm SER}=20\log\frac{{\rm RMS}\{\vet{z}\}}{{\rm RMS}\{\vet{z}-\vet{z}_M\}}.
\end{equation}
For any fixed \rm{SNR} level, we computed the time-domain \mbox{TD-SER} by averaging the performance induced by~\eqref{eq:SER} over the $R$ models and the two output signals. The frequency-domain \mbox{FD-SER} was computed in the same way, by averaging over the transfer matrix elements. These two metrics are shown in the top panel of Fig.~\ref{fig:Ex_1_NoisyInputOutput}.

The meaning of the SER being above or below the black line threshold is that \RTVF\ is either rejecting or amplifying the presence of the noise on the data, respectively. Since the \mbox{TD-SER} is always above the plane bisector (solid black line), \RTVF\ is able to partially reject the presence of measurements noise in the training data. This noise rejection property is expected, since the basis functions involved in the estimation procedure effectively filter the noisy input and output signals via~\eqref{eq:filtered_z}. On the other hand, \mbox{FD-SER} follows the bisector almost exactly, confirming also a good frequency-domain accuracy.

The noise-corrupted training input signals for a representative test case for ${\rm SNR}=16$ are depicted in the middle panel of Fig.~\ref{fig:Ex_1_NoisyInputOutput}, whereas the corresponding extracted model is validated against the reference time-domain output in the bottom panel. Even with this significant amount of noise, the time-domain prediction capabilities of the model are excellent.

\junk{
For fixed {\rm SNR} we also evaluated the impact of the noise on the frequency domain accuracy by averaging the following relative error {\color{TB} T: Now the frequency domain error is analogous to the SER. The figure has changed accordingly. Still I have to change the text following the equation...}
\begin{equation}\label{eq:freqErrorDef}
    E^r_{i,j}=20 \log \frac{ {\rm RMS} \{ \vet{\data{h}}_{i,j}  \} }     {  {\rm RMS} \{ \vet{\data{h}}^r_{i,j}- \vet{h}^r_{i,j}  \}},\quad 
    r=1,...,R, \quad i,j=1,2
\end{equation}
over all of the $R$ models and over all of the transfer function elements. Here $\vet{\data{h}}^r_{i,j}$ and $\vet{h}^r_{i,j}$ are vectors containing the samples of the $(i,j)$-th entry of the $r$-th reference system and model transfer matrices, respectively.
The results are reported in Fig.~\ref{fig:FreqDomainErrors}. The fitting error is within the $5\%$ accuracy up to ${\rm SNR}=30$. For higher noise power levels, the accuracy in the frequency domain is degraded, although the time domain fitting is still acceptably accurate, as shown in Fig.\ref{fig:Ex_1_NoisyInputOutput}.
}

\junk{
\begin{figure}
    \centering
    \includegraphics[width=0.9\columnwidth]{./Figures/SyntheticWithNoise/ErrorFRequency-eps-converted-to.pdf}
     \caption{The trend of the averaged frequency-domain relative error~\eqref{eq:freqErrorDef} as a function of the {\rm SNR}. The Error is within $5 \%$ for all of the responses greater than ${\rm SNR}=30$.}
    \label{fig:FreqDomainErrors}
\end{figure}
}

\subsection{Generator Modeling in the IEEE 39-Bus System}
\begin{figure}
    \centering
    \includegraphics[width=0.95\columnwidth]{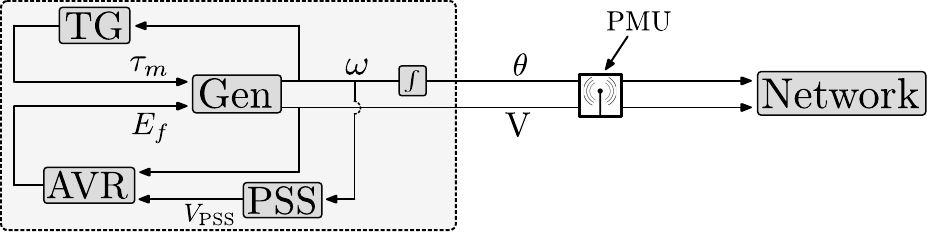}
     \caption{Shown is the interaction between the generator, its three controllers, and the network. The PMU collects data at the generator's point of connection. }
    \label{fig:GEN_Sys_BB}
\end{figure}

In order to test the performance of \RTVF\ in a simulated power system setting, we collected data from time domain simulations performed on the IEEE 39-Bus New England system. This system includes 10 generators and 19 ZIP loads; the associated load, network, and generator models and numerical parameters were taken directly from~\cite{hiskens:2013}. Accordingly, each generator was modeled as a $6^{\rm th}$ order synchronous machine with $3^{\rm rd}$ order automatic voltage regulators and power system stabilizers; additionally, each generator was outfitted with a $3^{\rm rd}$ order turbine governor (Type I)~\cite{Milano:2013}. Each generator system, with its three controllers, had a total of 15 dynamical states. The interaction between the generator, its controllers, and the network is shown in Fig.
\ref{fig:GEN_Sys_BB}. All time domain simulations were performed using MATLAB's DAE solver ode23t by setting relative and absolute error tolerances to $10^{-7}$ and $10^{-8}$, respectively.

\subsubsection*{Load Perturbations}
In order to mimic ambient load fluctuations, we applied an Ornstein-Uhlenbeck (OU) process~\cite{Milano:2013_OU} to the active and reactive power demands at each load. The dynamics of these processes are given by $\tau\dot{u}_{p} =- u_{p}+\eta_p$ and $\tau\dot{u}_{q} =- u_{q}+\eta_q$, where $\eta_p$, $\eta_q$ are zero-mean Gaussian variables, and the ``load reversal" time constant $\tau$ was set to 50s. We further applied a low-pass filter (LPF), with a cutoff frequency of $\sim\!\!9$Hz, to the OU variables such that $\hat{u}={\rm LPF}\{{u}\}$. This filtering operation was applied because dynamics above this frequency range become inconsistent with the quasi-stationary phasor approximation used in modeling the network's dynamics. Therefore, high frequency load behaviour is effectively neglected. Finally, these filtered OU variables were parameterized with time variable $t$ using a cubic spline interpolation and applied to the individual ZIP loads via
\begin{align}
P(t,{\rm V}) & =P_{0}(1+\beta\cdot\hat{u}_{p}(t))\bigl(a_{Z}\overline{{\rm V}}^{2}+a_{I}\overline{{\rm V}}^{1}+a_{P}\overline{{\rm V}}^{0}\bigr)\label{eq: P_noise}\\
Q(t,{\rm V}) & =Q_{0}(1+\beta\cdot\hat{u}_{q}(t))\bigl(b_{Z}\overline{{\rm V}}^{2}+b_{I}\overline{{\rm V}}^{1}+b_{P}\overline{{\rm V}}^{0}\bigr),\label{eq: Q_noise}
\end{align}
where $\overline{{\rm V}}\equiv{\rm V}/{\rm V}_0$. As a final step, scalar variable $\beta$ in (\ref{eq: P_noise})-(\ref{eq: Q_noise}) was experimentally tuned to a numerical value of $\beta=50$; this tuning generated load perturbations whose corresponding network voltage perturbations approximately matched those of real PMU data (in terms of signal strength).

We then applied \RTVF\ to data measured at the machine-network interface in order to model a single generator's closed-loop dynamics. For modeling purposes, we treated voltage magnitude ${\rm V}(t)$ and voltage phase $\theta(t)$ signals as inputs, and we treated current magnitude ${\rm I}(t)$ and current phase $\phi(t)$ signals as outputs. The \RTVF\ algorithm sought to generate a MIMO model with $P=2$ and various reduced orders $N$.

\begin{figure}
    \centering
    \includegraphics[width=0.9\columnwidth]{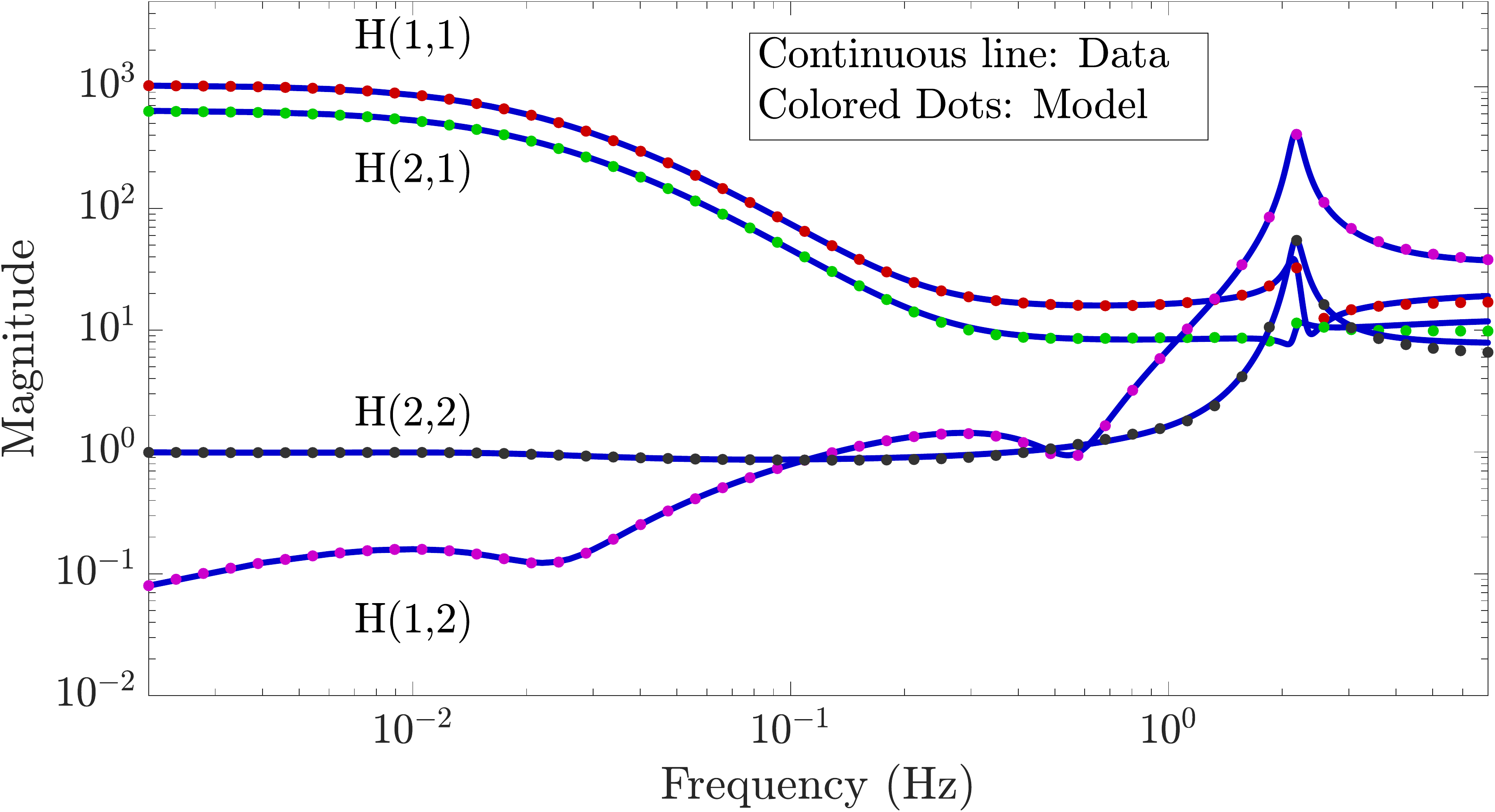}\\[2mm]
    \includegraphics[width=0.9\columnwidth]{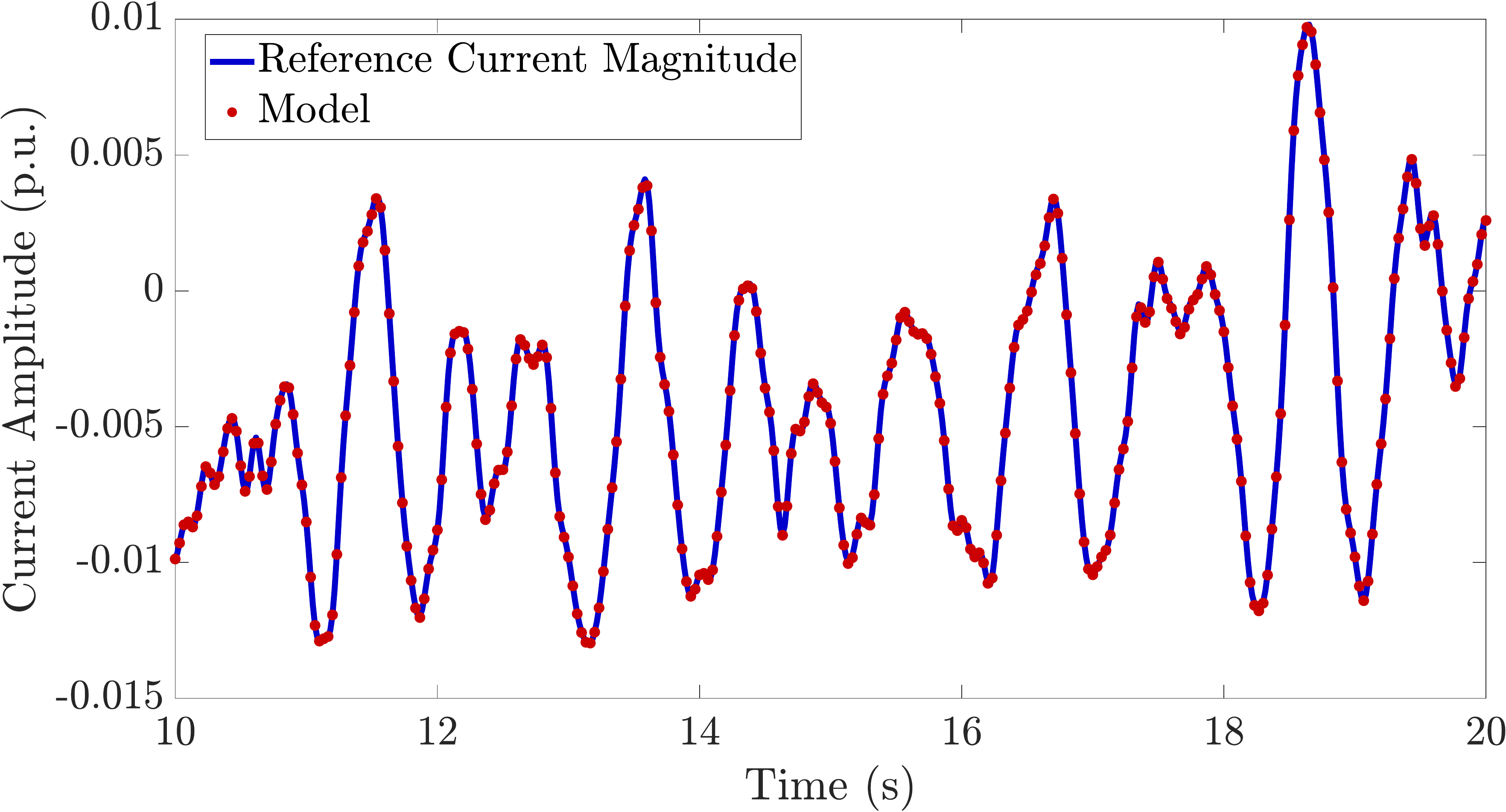}
    \vspace{0.3 cm}
     \caption{Noise-free generator model extraction. Top panel: frequency domain validation of the generator model (order $N=9$) against exact machine equations. Bottom panel: the time domain validation of the model against the current magnitude reference output.}
    \label{fig:GeneratorValidation}
\end{figure}

In order to validate the quality of the model generated in the absence of measurement noise, we refer to the time and frequency domain references provided by the exact machine equations. The results are provided in Fig.~\ref{fig:GeneratorValidation}. The results show the model accuracy is excellent in both the time and frequency domains, even though the reduced model order (in this case $N=9$) is less than the machine's true model order (${\bar N}=15$).

\begin{algorithm}
		\caption{Measurement Noise Application}\label{al:SNR}
		\small{\begin{algorithmic}[1]
			\INPUT Voltage \& current signals ${\rm V}(t)$, ${\rm I}(t)$, ${\theta}(t)$, ${\phi}(t)$; desired SNR
			\OUTPUT Noisy voltage \& current signals ${\rm V}_n(t)$, ${\rm I}_n(t)$, $\theta_n(t)$, $\phi_n(t)$
            \STATE $\sigma_{{\rm V}n} \leftarrow{\rm RMS}\left\{ {\rm V}(t)-{\rm E}\left\{ {\rm V}(t)\right\} \right\}\cdot10^{-{\rm SNR}/20}$
            \STATE $\sigma_{{\rm I}n} \leftarrow{\rm RMS}\left\{ {\rm I}(t)-{\rm E}\left\{ {\rm I}(t)\right\} \right\}\cdot10^{-{\rm SNR}/20}$
            \STATE ${\rm V}_{n}(t) \leftarrow{\rm V}(t)+\sigma_{{\rm V}n}\cdot\eta(t)$
            \STATE ${\rm I}_{n}(t) \leftarrow{\rm I}(t)+\sigma_{{\rm I}n}\cdot\eta(t)$
            \STATE $\theta_{n}(t)\leftarrow\theta(t)+\sigma_{{\rm V}n}\cdot\eta(t)/{\rm E}\left\{ {\rm V}(t)\right\}$
            \STATE $\phi_{n}(t) \leftarrow\phi(t)+\sigma_{{\rm I}n}\cdot\eta(t)/{\rm E}\left\{ {\rm I}(t)\right\}$
\junk{            \STATE $P_{n}(t) ={\rm V}_{n}(t){\rm I}_{n}(t)\cos(\theta_{n}(t)-\phi_{n}(t))$
            \STATE $Q_{n}(t) ={\rm V}_{n}(t){\rm I}_{n}(t)\sin(\theta_{n}(t)-\phi_{n}(t))$}
            \RETURN ${\rm V}_n(t)$, ${\rm I}_n(t)$, $\theta_n(t)$, $\phi_n(t)$
		\end{algorithmic}}
\end{algorithm} 

\subsubsection*{Measurement Noise}
In order to further gauge the practical effectiveness of \RTVF, we applied measurement noise to the voltage and current signals measured by the PMU. To apply this noise, we utilized the procedure outlined in Algorithm \ref{al:SNR}, where $\eta(t)$ represents an AWG noise vector. In this algorithm, an SNR is first specified in terms of magnitude ($\rm V$, $\rm I$) signals\junk{\footnote{Because phase angles drift aggressively, faithfully defining their ``signal strength" for noise application purposes is not a straightforward procedure.}}, and the proper amount of noise is then added. Next, noise with an appropriate standard deviation is applied to the phase signals ($\theta$, $\phi$), such that the total vector error (TVE) in the complex plane would be a circular cloud. In other words, an ``equivalent" amount of noise is applied to both the magnitude and phase data, relative to the specified SNR value.
\junk{Finally, the noisy voltage and current data are used to compute noisy active $P_n(t)$ and reactive $Q_n(t)$ power signals.}

\begin{figure}
    \centering
    \includegraphics[width=0.9\columnwidth]{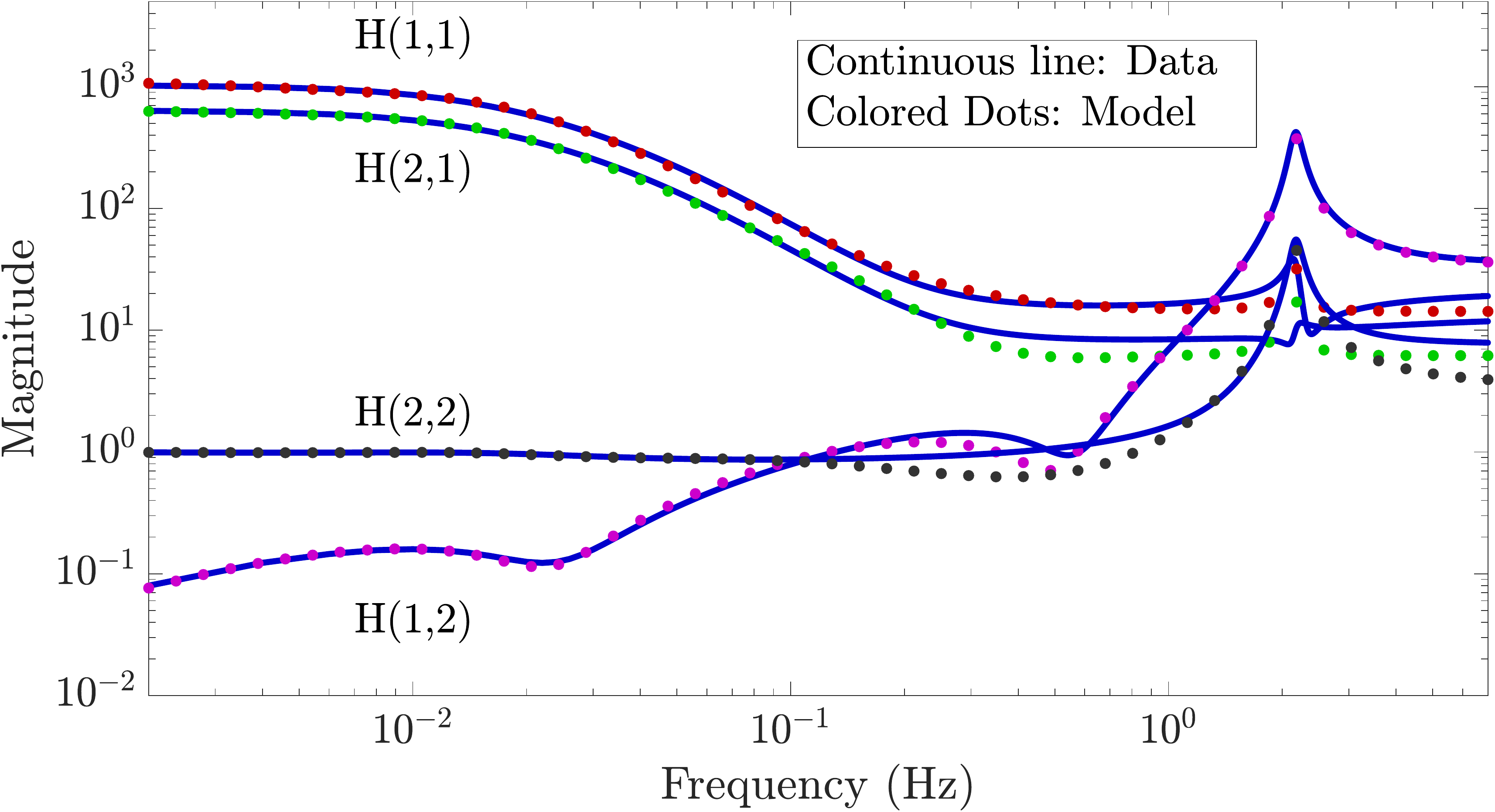}\\[2mm]
    \includegraphics[width=0.9\columnwidth]{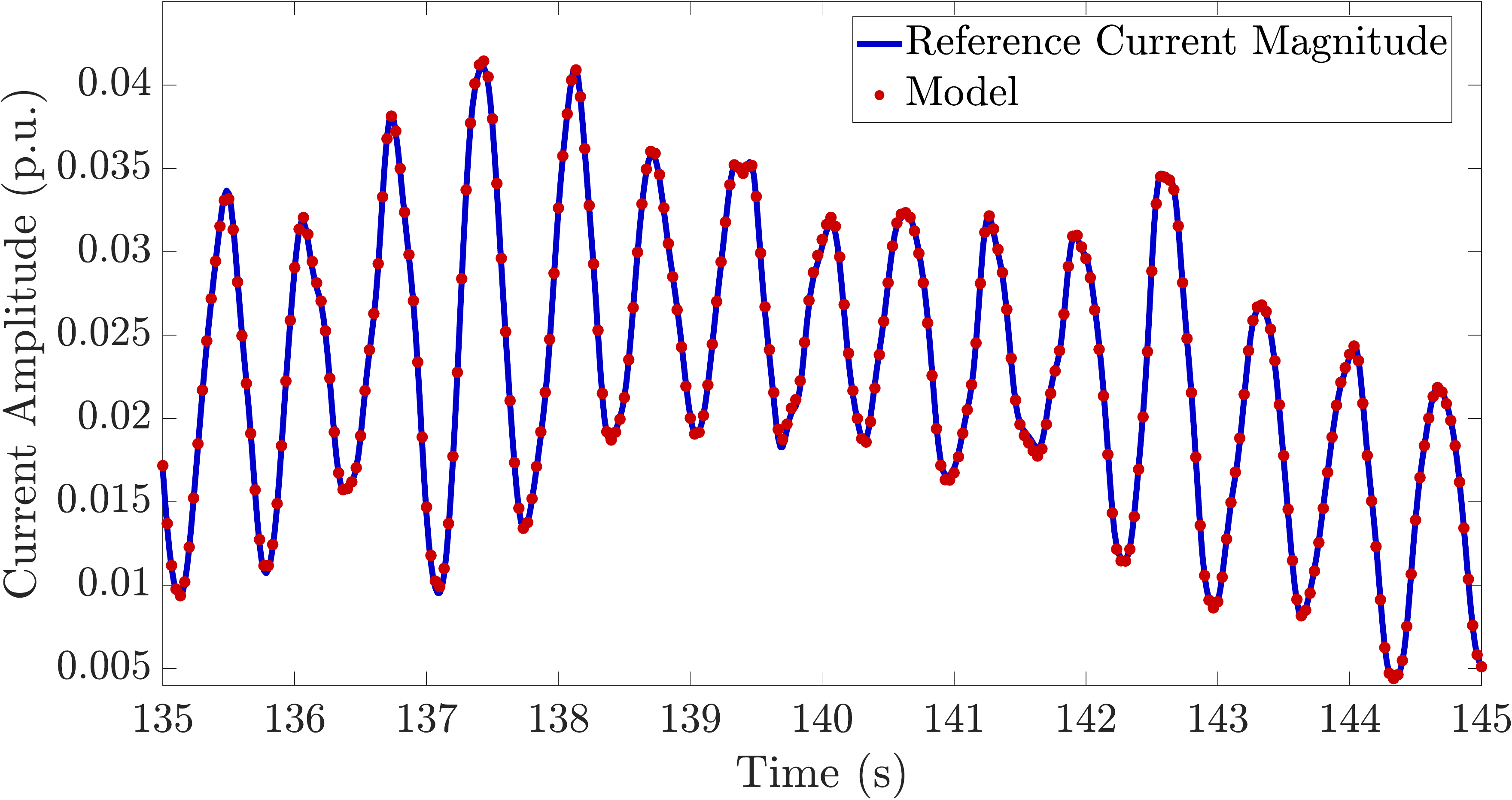}\\[2mm]
    \includegraphics[width=0.9\columnwidth]{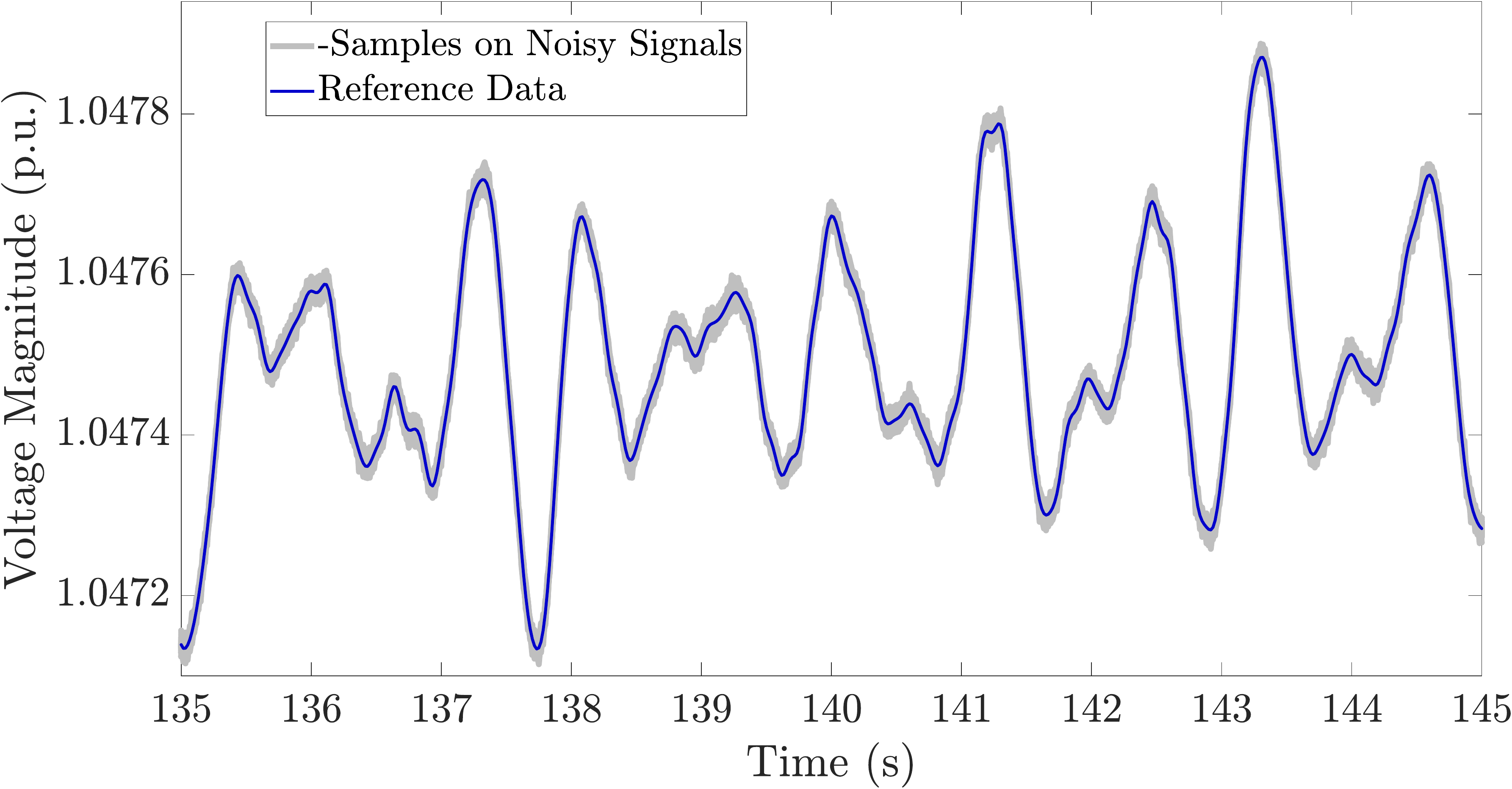}
     \caption{Generator model training from noisy data (\mbox{${\rm SNR}=32$}). Top panel: frequency responses. Middle panel: small signal current magnitude. Bottom panel: noise-corrupted input training signals (samples from a cloud of $R=100$ different realizations depicted with a shade of grey).}
    \label{fig:SNRoutputGenerator}
\end{figure}

Top and middle panels of Fig.~\ref{fig:SNRoutputGenerator} report the frequency- and time-domain fitting performance of a \RTVF\ model of order $N=7$ obtained for ${\rm SNR}=32$dB. The corresponding noise-corrupted voltage magnitude signals are depicted in the bottom panel. Compared to the performance in the noise-free setting (Fig.~\ref{fig:GeneratorValidation}), these results show that the frequency-domain model accuracy is still quite acceptable, and that the accuracy in the time domain seems to be not affected by the presence of noise. Therefore, we conclude that the time prediction capabilities of \RTVF\ models extracted from noisy signals are potentially adequate for power system applications.

\subsection{Wide Area Monitoring in the IEEE 39-Bus System} \label{subsec:line}
\begin{figure}
    \centering
    \includegraphics[viewport=57bp 90bp 720bp 515bp,clip,width=1\columnwidth]{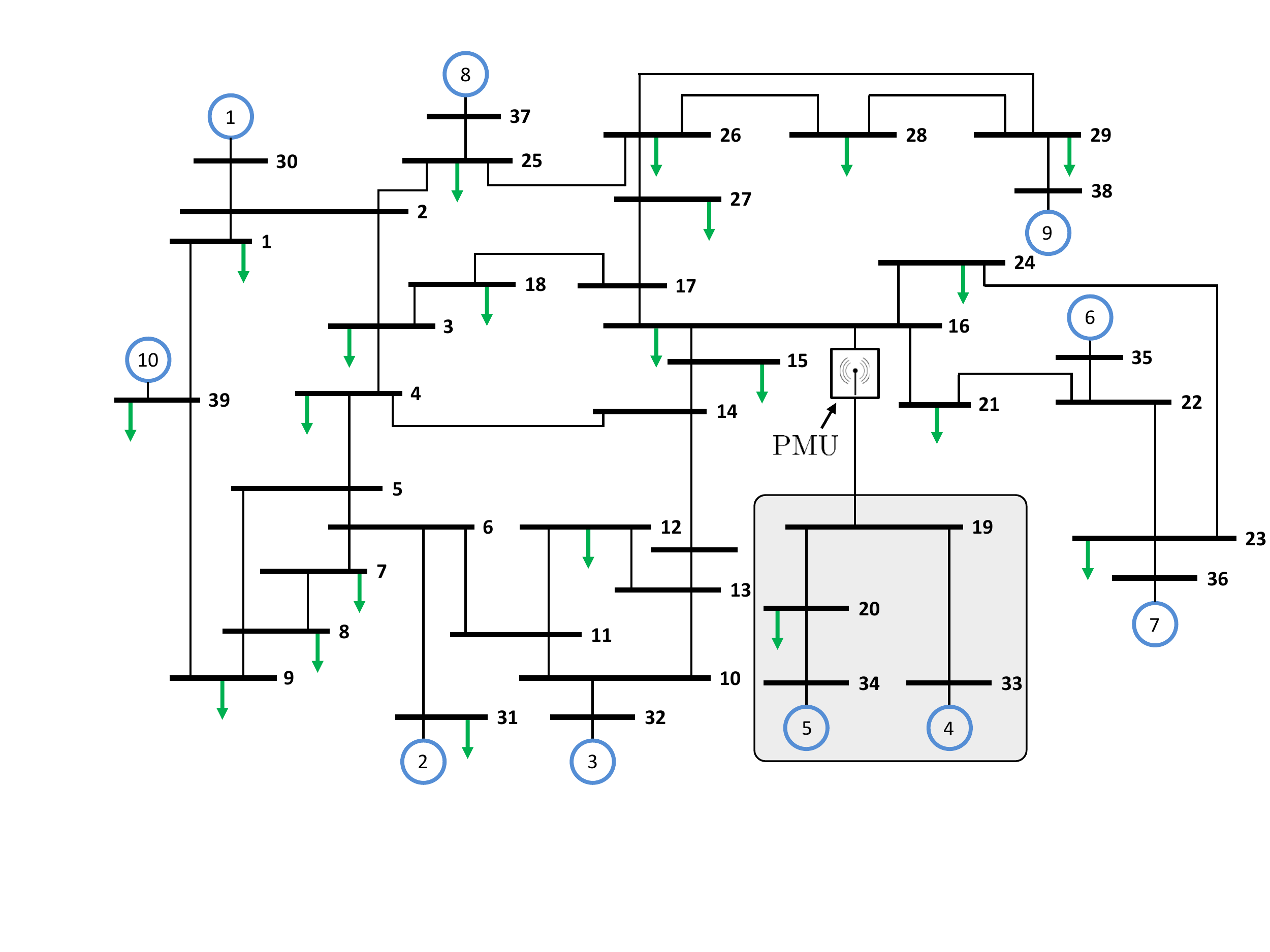}
     \caption{IEEE 39-bus New England system. The area depicted by the grey box, which geographically corresponds to a region in southern Massachusetts, US, is used to test \RTVF's ability to perform wide area predictive modeling.}
    \label{fig:39_Bus}
\end{figure}

Finally, as a third experiment, we tested \RTVF's ability to perform ``wide area'' monitoring in the 39-Bus system via predictive modeling. {The proposed approach is here compared with the ARX modeling~\cite{Chai:2016,stoica2005spectral} and the standard Time Domain Vector Fitting scheme applied without the inclusion of initial conditions estimation}. To do so, we measured the simulated current flowing across line 16-19, as shown in Fig. \ref{fig:39_Bus}, along with the voltage perturbations on bus 16. Subsequently, we used the resulting time domain data (i.e., ${\rm V}(t)$, $\theta(t)$, ${\rm I(t)}$, $\phi(t)$) to model the linearized dynamics of the grey box depicted in Fig. \ref{fig:39_Bus}, consisting of two generators, a load\footnote{The stochastic perturbations at this load were assumed negligible, such that the observed dynamics were fully deterministic.}, and their interconnecting lines. {We collected the data at a $60$ Hz sampling rate for a total duration of $500$ seconds, using the training subset  $t\in[100,360]~\rm{s}$ to generate the models using the three considered methods.} The dynamics of this wide area had $\bar N =30$ full order states, but using the ambient PMU data, we were able to derive a reduced order model of dynamic order $N=13$.

\begin{figure}
    \centering
    \includegraphics[width=\columnwidth]{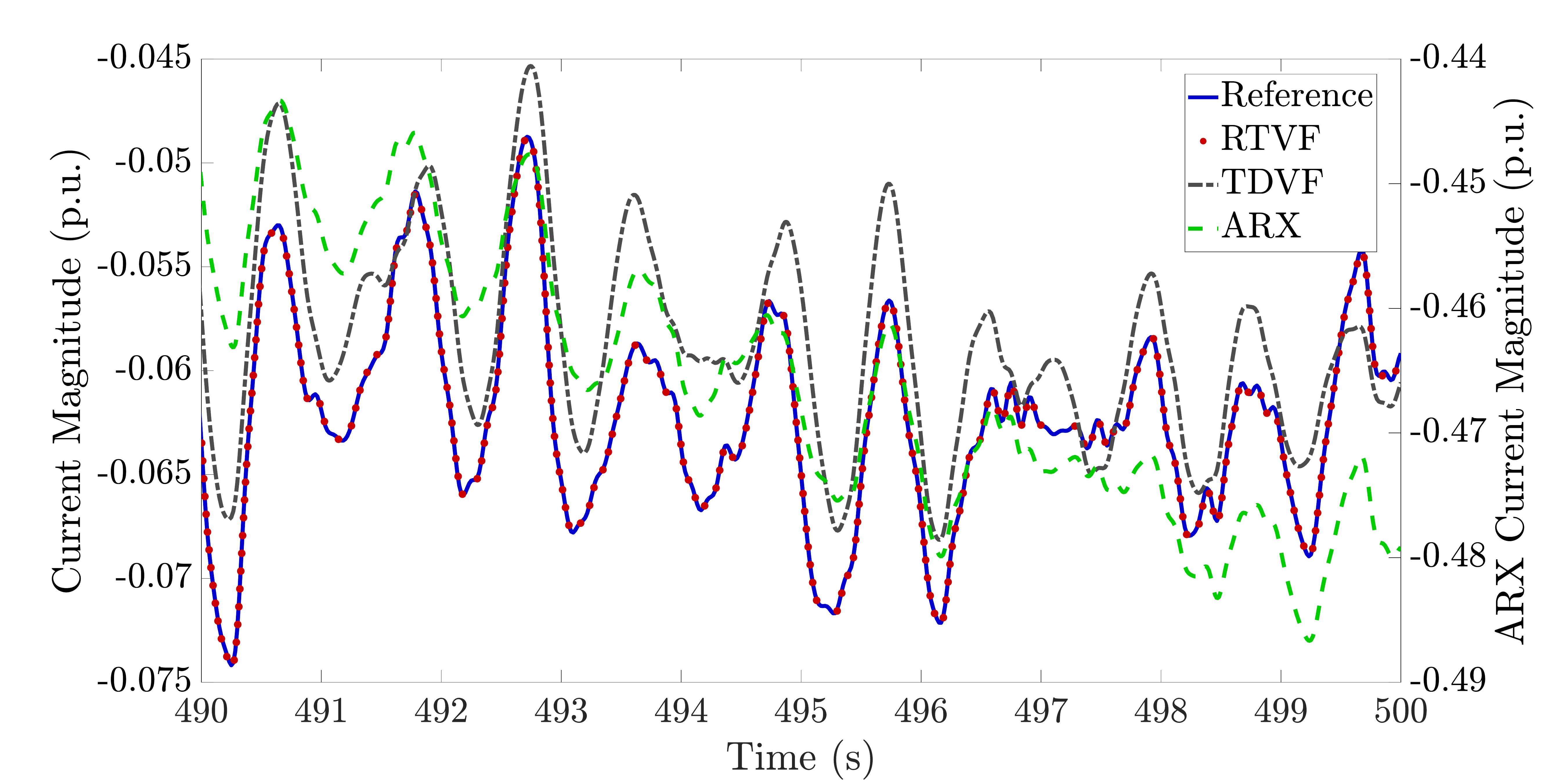}\\[2mm]
    \includegraphics[width=\columnwidth]{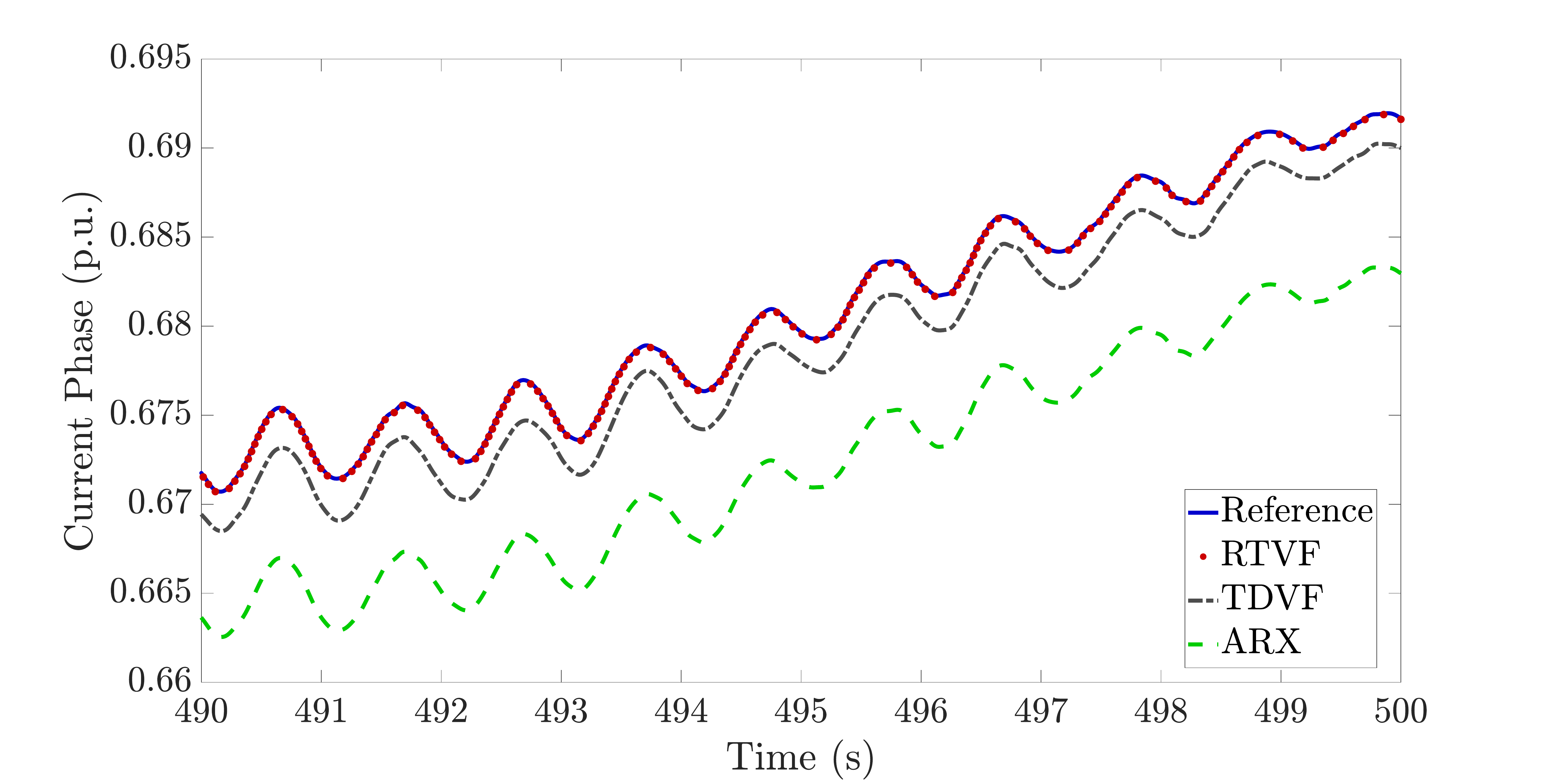}\\[2mm]
     \caption{Wide area test case. Time domain validation of models derived by means of RTVF, ARX, and TDVF without inclusion of the initial conditions estimate. Top panel: Current magnitude deviation (note that the left y-axis applies to RTVF and TDVF, while right y-axis refers to ARX; the latter has difficulties in producing a sound approximation, especially at low frequencies). Bottom panel: Current phase deviation. }
    \label{fig:LineOutput}
\end{figure}

\begin{figure}
    \centering
    \includegraphics[width=\columnwidth]{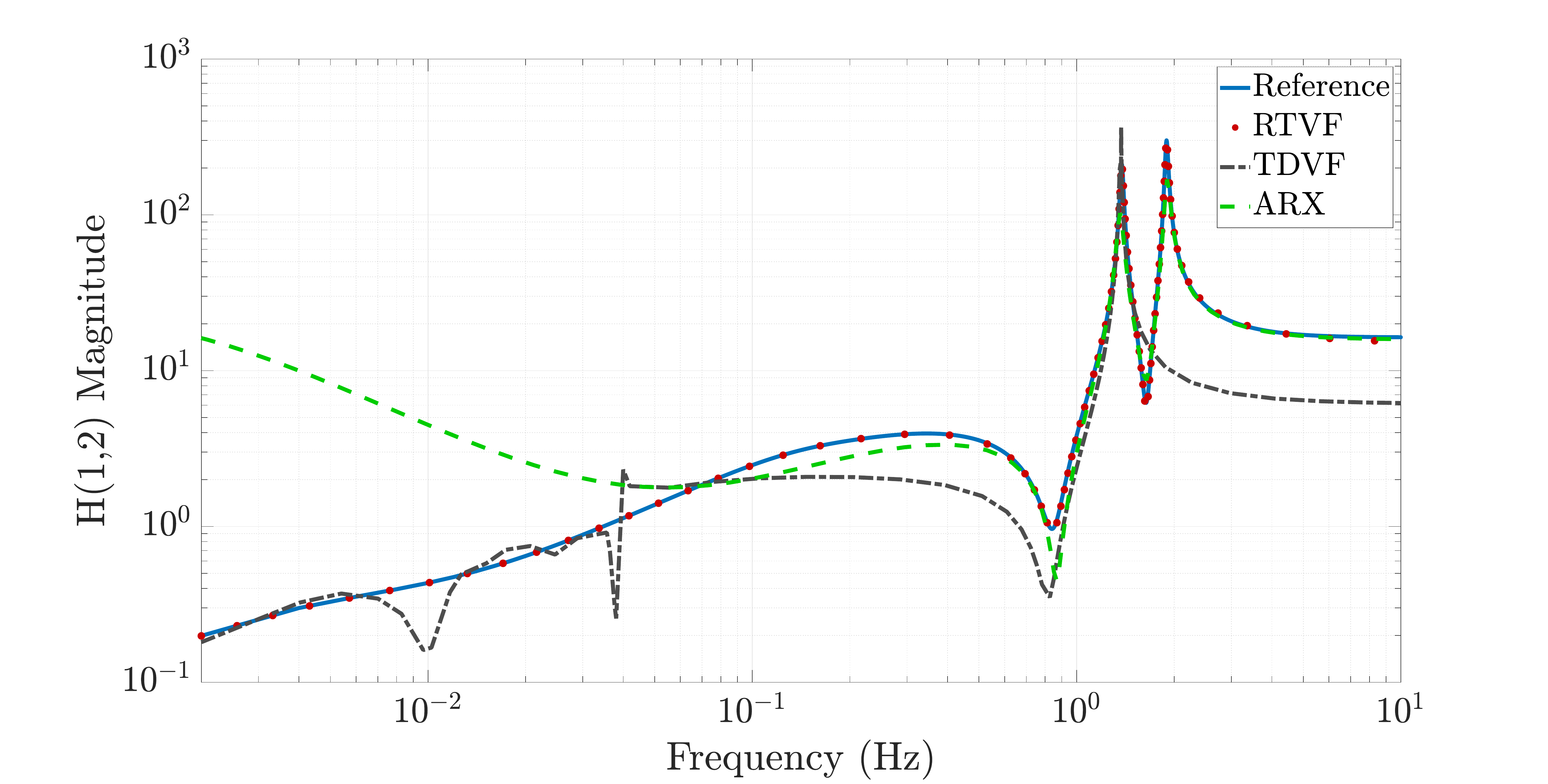}\\[2mm]
     \includegraphics[width=\columnwidth]{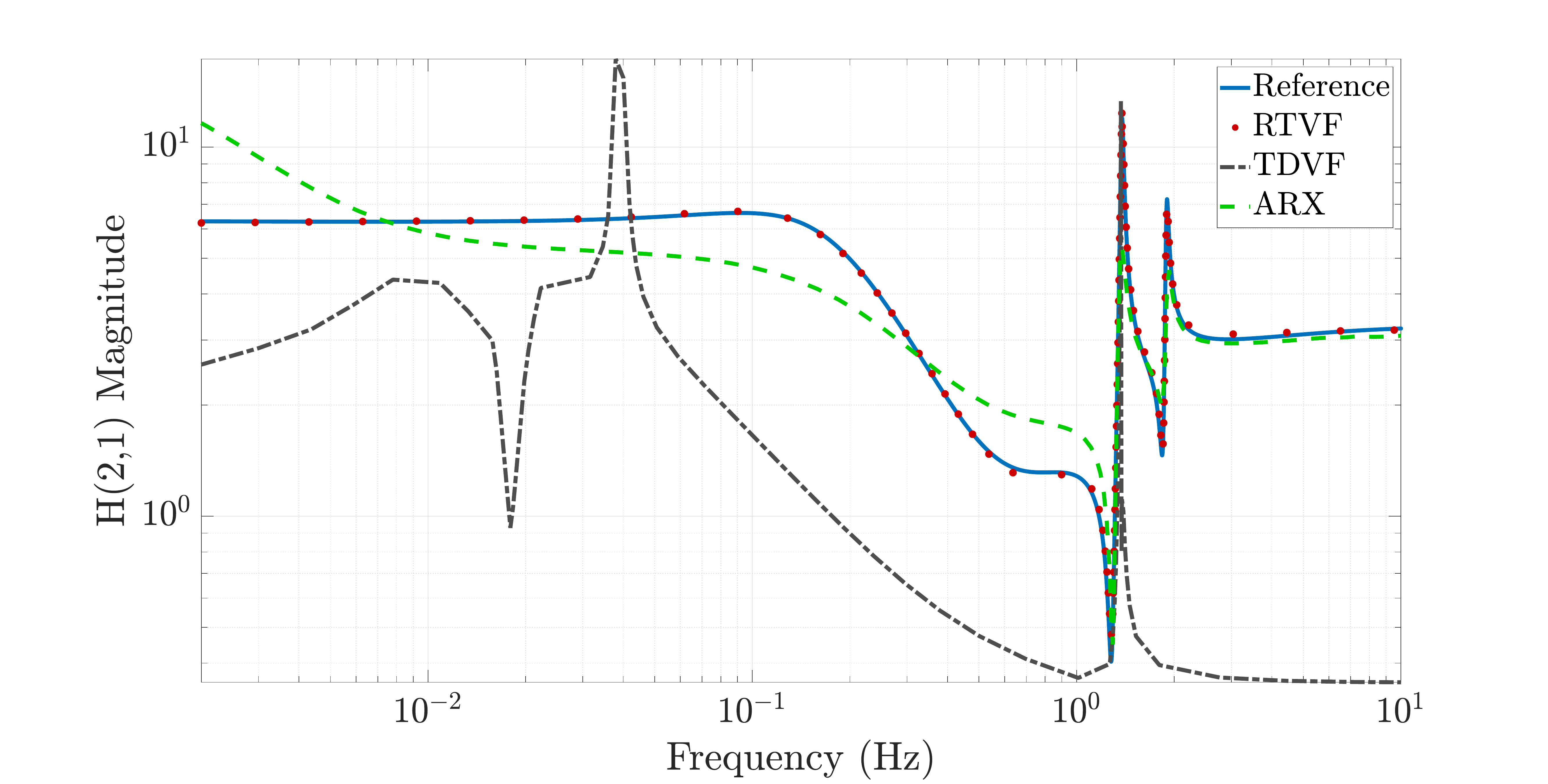}\\[2mm]
     \caption{Selected transfer functions elements estimated by the \RTVF, \TDVF~and ARX models, compared to the reference (exact) responses.}
    \label{fig:LineTFs}
\end{figure}

{The results are shown in Fig.~\ref{fig:LineOutput}, where we compare the accuracy of the models in terms of output current magnitude and phase (time domain). The results we provide are referred to time domain validation data, that were not used for model generation, and that are therefore meaningful to validate the models' respective qualities; for the sake of visualization, we used a different y-axis scale for the current magnitude signal returned by the ARX model (right y-axis in the figure top panel). We observe that the proposed \RTVF\ method outperforms ARX and \TDVF\ algorithms, providing highly accurate predictions. We remark that, in principle, the application of \TDVF\ in its standard formulation is not even conceivable in the considered scenario due to the presence of concurrent inputs, which were here handled by means of the proposed modified scheme of Sec.~\ref{sec:Implementation}. Nevertheless, the experiments confirm that including the effect of the initial conditions is crucial to deriving a meaningful model. This is even more evident when the frequency responses of the models are compared with those of the reference system, as reported in Fig.~\ref{fig:LineTFs} for the two off-diagonal elements of the transfer matrix; the remarkable accuracy of the \RTVF\ model is not achieved by the ARX and \TDVF\ algorithms.}

In the bottom panel, we showcase the model's excellent frequency domain accuracy by comparing it against the two off-diagonal elements of the reference transfer function. \RTVF's capacity for accurately modeling aggregated power system dynamics, even at the wide area level, is clearly demonstrated by the high-fidelity results in Fig.~\ref{fig:LineOutput},~\ref{fig:LineTFs}.

\junk{
\begin{figure}
    \centering
    \includegraphics[width=\columnwidth]{./Figures/IEEE39/H11-eps-converted-to.pdf}\\[2mm]
     \caption{H11}
    \label{fig:LineH11}
\end{figure}
}
\junk{
\begin{figure}
    \centering
    \includegraphics[width=\columnwidth]{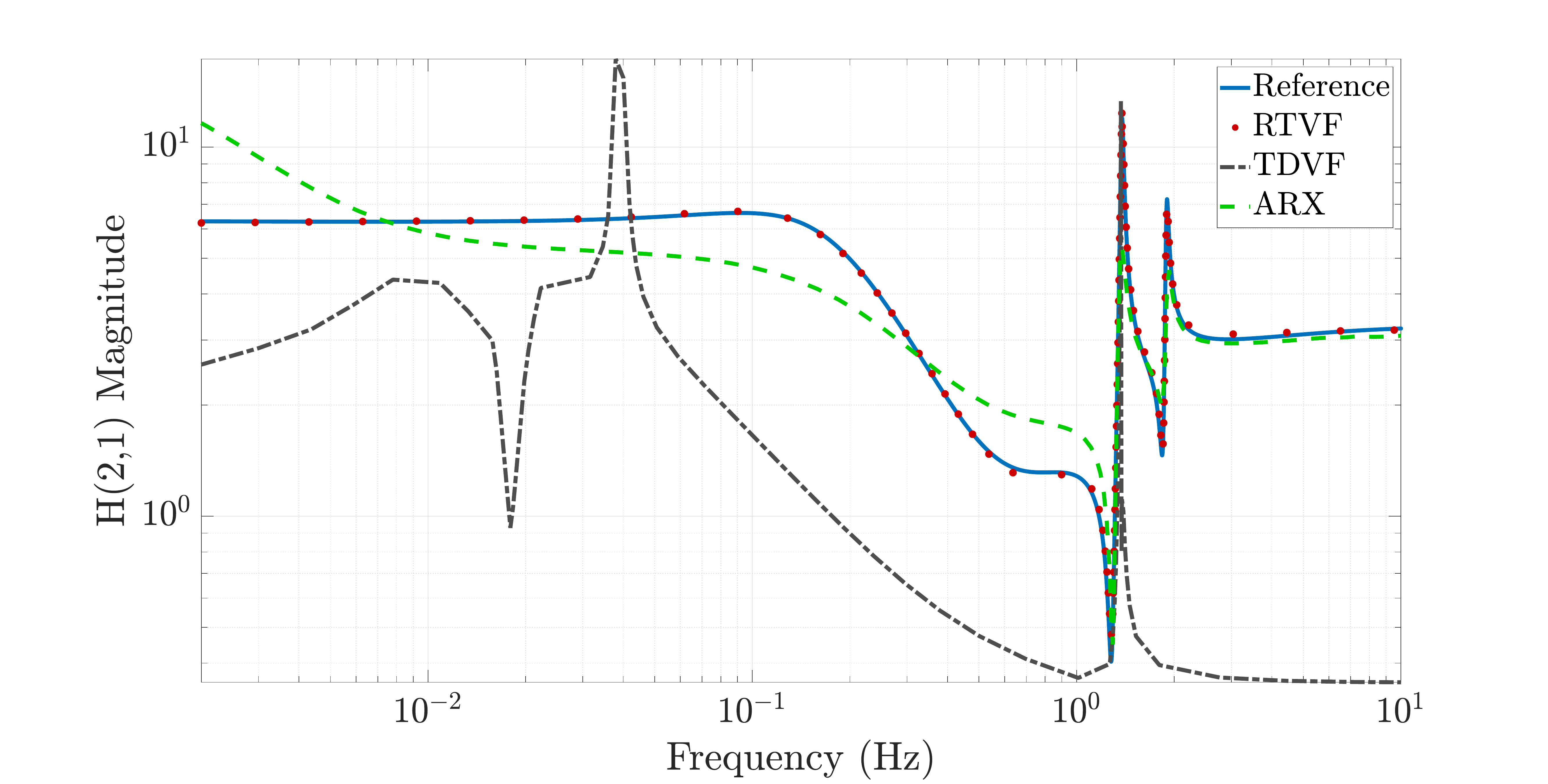}\\[2mm]
     \caption{H21}
    \label{fig:LineH21}
\end{figure}
\begin{figure}
    \centering
    \includegraphics[width=\columnwidth]{./Figures/IEEE39/H22-eps-converted-to.pdf}\\[2mm]
     \caption{H22}
    \label{fig:LineH22}
\end{figure}
}


\section{Conclusions}
\label{sec:Conclusions}

This work proposed a new procedure, termed Real-Time Vector Fitting (\RTVF), which performs real-time predictive modeling of linearized dynamics. While the \RTVF\ algorithm can be applied across a broad range of dynamical engineering systems, it was specifically developed to perform real-time predictive modeling of power system dynamics in the presence of ambient perturbations. Accordingly, \RTVF\ {has the following features, both of which are necessary for practical predictive modeling of power systems: 
\begin{itemize}
\item It explicitly accounts for the presence of initial state decay in the observed output of a power system component (e.g., generator, load, or wide-area). 
\item It can be applied in the face of concurrently active input signals, since this is a practical constraint for any three-phase power system component modeled using a quasi-stationary phasor approximation at its terminals.
\end{itemize}
Furthermore, we compared the performance of \RTVF\ to that of the canonical \TDVF\ and ARX algorithms. In our experiments, the RMS error committed by \RTVF\ on a single transfer function element was found to be about one order of magnitude smaller than what \TDVF\ and ARX allowed for. The latter methods provide very poor prediction at lower frequencies due to an incorrect treatment of initial state decay contributions, whereas \RTVF\ was demonstrated to produce extremely accurate predictions over a broad frequency band.}


Given the widescale deployment of PMUs and the emerging stability concerns associated with low-inertia grids, \RTVF\ provides power system operators with a valuable tool. This tool, for example, can build real-time models of power system components whose physical prior models are unknown. It can also validate or enhance prior models which are supposedly well known. Additionally, since controller parameters (e.g., droop gain) are often changed at the local level without system operator awareness, extensions of the \RTVF\ scheme can be used to infer the true value of these parameters in real-time, as was analogously demonstrated in~\cite{Jakobsen:2017}. Such capability can allow grid operators to maintain small-signal stability and ensure that generators are properly adhering to market regulations. To enhance its practical effectiveness, future research will couple \RTVF\ with advanced noise filtering mechanisms. Also, physically-aware regularization techniques will be developed to allow operators to more effectively ``track'' a system's shifting dynamics as operating equilibriums evolve over time.


\bibliographystyle{IEEEtran}
\bibliography{RTVF_Bib,PM_Biblio}

\end{document}